\documentclass[twocolumn,amsmath,amssymb,longbibliography]{revtex4-2}

\usepackage{xcolor}
\usepackage{graphicx}
\usepackage{subfigure}
\usepackage{dcolumn}
\usepackage{bm}
\usepackage{framed}
\usepackage{hyperref}
\usepackage{dsfont}
\usepackage[normalem]{ulem}
\usepackage{nicefrac}
\usepackage{braket}
\newcommand{\el}[1]{\textcolor{black}{#1}}
\newcommand{\er}[1]{\textcolor{black}{#1}}
\newcommand{\cor}[1]{\textcolor{black}{#1}}
\newcommand{\core}[1]{\textcolor{black}{#1}}
 \begin{document}
	
\title{\cor{A many-\core{particle} bosonic} quantum  Maxwell demon}

\author{Josef Hlou\v sek}
\affiliation{Department of Optics, Palack\'y University, 17. listopadu 1192/12, 771 46, Olomouc, Czech Republic}
\author{Tobias Denzler}
\affiliation{Institute for Theoretical Physics I, University of Stuttgart, D-70550 Stuttgart, Germany}
\author{Vojt\v ech \v Svarc}
\affiliation{Department of Optics, Palack\'y University, 17. listopadu 1192/12, 771 46, Olomouc, Czech Republic}
\author{Miroslav Je\v zek}
\affiliation{Department of Optics, Palack\'y University, 17. listopadu 1192/12, 771 46, Olomouc, Czech Republic}
\author{Eric Lutz}
\affiliation{Institute for Theoretical Physics I, University of Stuttgart, D-70550 Stuttgart, Germany}
\author{Radim Filip}
\affiliation{Department of Optics, Palack\'y University, 17. listopadu 1192/12, 771 46, Olomouc, Czech Republic}

\begin{abstract}
Energy extraction from a measured quantum system is a cornerstone of information thermodynamics as illustrated by Maxwell's demon. The nonequilibrium physics of many-particle systems is additionally  strongly influenced by quantum statistics. We here report the first experimental realization of a quantum demon in a many-particle photonic setup made of two identical thermal light beams. We show that single-photon measurements combined with feedforward operation may deterministically increase the mean energy of one beam faster than energy fluctuations, thus improving the thermodynamic stability of the device. We moreover demonstrate that bosonic statistics can enhance the energy output above the classical limit, and further analyze the counterintuitive thermodynamics of the demon using an information-theoretic approach. Our results underscore the pivotal role of many-particle statistics for enhanced energy extraction  in quantum thermodynamics.
\end{abstract}

\maketitle
Maxwell's   demon reveals a profound relation between information and thermodynamics.
By measuring    positions and velocities of  gas particles contained in two neighboring chambers connected by a small aperture that can be opened or closed with a trapdoor, the demon exploits the acquired information to collect fast (hot) particles in one chamber and slow (cold) particles in the other \cite{lef03,ple01,mar09}. The resulting temperature difference may then be used   to run a heat engine and perform work by lifting a weight---without the demon investing any work himself, in apparent violation of the second law \cite{par15,lut15}.  Originally a thought experiment, \core{realizations of demonic devices have lately} been reported in a  number of classical \cite{rai09,toy10,rol14,kos14,kos14a,kos15,chi17,kum18,pan18,adm18,rib19,deb20,sah21,sah23,yan24} and single-particle quantum \cite{cam16,cot17,mas18,nag18,naj20} systems. 

\cor{A basic property of identical quantum particles is indistinguishability. Whereas classical particles are distinguishable, and described by Maxwell-Boltzmann statistics, identical quantum particles are either bosonic or fermionic depending on the symmetry or antisymmetry of their  state under permutation \cite{aul09}. The role of quantum statistics in the operation of a bosonic Szilard information engine  has been studied theoretically and  found to enhance the work output owing to bunching-related effects \cite{kim11,cai12,jeo16,ben18,ben18a} (see also discussions in Refs.~\cite{ple13,kim13,ple14}). Similar bosonic boosts have been predicted in an optomechanical piston \cite{hol20} and in many-particle  heat engines \cite{mye20,wat20}. Quantum  demon experiments  have so far  been realized  with single (fermionic) qubit systems \cite{cam16,cot17,mas18,nag18,naj20}.  In addition, photonic demons have been implemented in the ultra-low light  limit, with a mean photon number much smaller than one \cite{zan22}, and in the  intense light  limit,  with  a mean photon number of the order of $10^8$ \cite{vid16}. The \core{first experiment} corresponds to the single particle limit, while the \core{second} to classical statistics. To our knowledge,  the influence of quantum statistics on a many-\core{particle} demon has not been explored experimentally yet.}

  \begin{figure*}[ht!]
	\centering
	\includegraphics[width=1.0\linewidth]{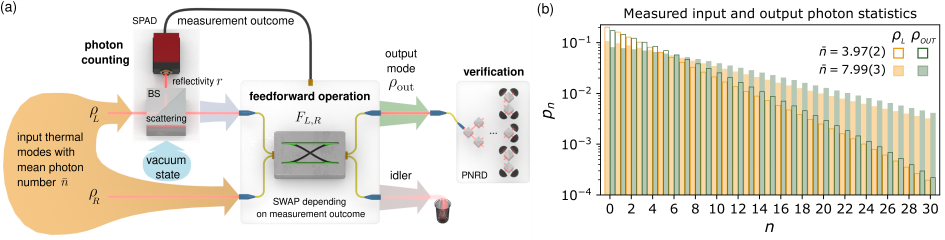}
	\caption{\el{(a) \core{The experimental setup consists of two input thermal light modes L and R with mean photon number $\bar n$. A tunable-reflectivity beam splitter (BS) weakly measures mode L by coherently scattering a varying fraction r of photons into vacuum modes. The scattered photons are detected with single-photon avalanche diode (SPAD), and the measurement result is used to perform an optical SWAP operation that flips either mode L or R to the output mode $\rho_{out}$, depending on the measurement outcome. This ensures that the energy of the output mode is deterministically higher than that of the input mode.} The photon statistics of the \core{optical} state $\rho_\text{out}$ is measured with a photon-number-resolving detector (PNRD). b) \cor{Measured photon statistics for the input ($\rho_L$) and output ($\rho_\text{out}$) modes for two values of the mean photon number $\bar n$, \core{for a scattering ratio $r = 14.4(1)\%$}}.
	}}
	\label{fig_1}
\end{figure*}

We here present the realization of a quantum photonic demon by employing two identical thermal light beams with a small mean photon number per mode \el{($\bar{n}=3.97(2)$ and $7.99(3)$)}. In this microscopic regime, discrete photon fluctuations  \cor{and quantum statistics} dominate. We combine weak single-photon measurements akin to photon substraction techniques \cite{nee06,our06,par07,Zavatta2008,Fedorov2015,Bogdanov2017,Hlousek2017,HashemiRafsanjani2017,Katamadze2018} on the first beam together with active  feedforward operation, that may swap the two light beams depending on the measurement outcome, to \el{deterministically} increase the mean energy of the beam. We  examine in detail the average extracted energy, the extracted energy fluctuations, the corresponding mean-to-deviation ratio as well as the second-order photon autocorrelation function of the output mode  \cite{Loudon2010}. We show that the  measurement-feedforward protocol effectively reduces photon correlations and energy fluctuations. At the same time,  we find that it  is able to  enhance  the mean energy as well as the mean-to-deviation ratio, that is, the stability \cite{pie18,hol18,den21}, of the output mode  above the thermal limit. \cor{We further   analyze the effects of bosonic quantum statistics, and demonstrate that it may enhance the mean energy output compared to the classical limit. Finally, we investigate the nonequilibrium thermodynamics of the quantum demon using an  information-theoric approach, and find that it satisfies a generalized (reversed) Clausius equality  \cite{par89,per02,gav14,may23}, \core{in contrast to a classical Maxwell demon}.}

 \textit{Experimental setup.}    Our system consists of two  (input) thermal light modes at inverse temperature $\beta$  with density operator  $\rho_{L,R}= \sum_n \exp{(-\beta\hbar\omega n)}|n\rangle\langle n|/(\bar n+1)$,  mean photon number $\bar n= 1/(\exp(\beta\hbar\omega-1)\el{)}$ and standard deviation $\bar{\sigma}=\sqrt{\bar{n}(1+\bar{n})}$, where $\omega$ denotes the optical frequency. The two modes play the role of the left (L) and right (R) chambers in Maxwell's thought experiment. Information about the system is gained by coherently scattering a fraction $r$ of the photons of mode L into empty vacuum modes and measuring $m$ scattered photons with a single-photon detector (mode R is not measured) (Fig.~1a). To that end, we employ a tunable-reflectivity beam splitter realized with a half-wave plate followed by a polarizing beam splitter.  The weak measurement is optimized by varying the reflectivity $r$. \cor{Such quantum-optical systems are advantageous from a thermodynamic point of view since their unbounded energy spectra allow them to accumulate large numbers of quanta per single mode, in contrast to systems with finite Hilbert spaces \cite{cam16,cot17,mas18,nag18,naj20}. Moreover, {surrounding} empty modes naturally act as a zero-temperature quantum environment, without the need of any external cooling  \cite{Loudon2010}}.
  
 The probability to detect $m$ photons is $p(m)=\text{Tr}[M(m) S(r) {\rho}_L S^\dagger(r) M^\dagger(m)]$, where  $S(r)$ is a unitary beam splitter-like transformation  that characterizes the  scattering process and $M(m)$ is a positive operator-valued measure (POVM) that describes the  photodetection process (see Supplemental Material for details \cite{sup}). The measurement  affects the photon distribution, which becomes nonthermal \cite{nee06,our06,par07}.  When a photon is detected, the average energy of mode L is enhanced compared to that of mode R; it is \el{otherwise} reduced when no photon is detected. In the limit $r\ll1$, such photon substraction \el{protocol}  is known  to conditionally increase the mean photon number of thermal light up to $(m+1)\bar{n}$ \cite{nee06,our06,par07,Zavatta2008,Fedorov2015,Bogdanov2017,Hlousek2017,HashemiRafsanjani2017,Katamadze2018}.
 We next implement  feedforward control by swapping the two modes L and R, depending on the measurement outcome, to ensure that the output  always has an energy larger than the \el{input mode} L. We concretely use a tunable photon routing device \cite{Svarc2019} to realize the   operation
 \begin{equation}
F_{L,R}(m) = 
\begin{cases}
\text{SWAP}_{L,\text{out}} &\text{if } m \geq 1 \\
\text{SWAP}_{R,\text{out}} &\text{if }  m  < 1,
\end{cases}
\end{equation} 
where 'out' denotes the output mode $\rho_\text{out}$.

 \textit{\core{Energetics, stability, and quantum statistics.}} We begin by analyzing the energetics and \core{stability}  of the output mode for  two different input mean photon numbers, \el{$\bar{n}=3.97(2)$ and $7.99(3)$}. We measure the click statistics of $\rho_\text{out}$ with a photon-number-resolving detector made of a balanced \el{ten}-channel spatially multiplexed optical network and \el{ten} single-photon avalanche diodes. \cor{From this multiple-coincidence measurement, we \core{fully} reconstruct the photon statistics $p_n$ using a retrieval algorithm based on entropy regularization of the maximum-likelihood estimation \cite{Hlousek2019}, as shown in Fig.~1b}. The energy extracted by the measurement-feedforward protocol is equal to $\langle \Delta E \rangle= \hbar\omega (\langle n\rangle - \bar n)$, where $\langle n\rangle$ is the average photon number of the output mode. The corresponding energy fluctuations are $\langle \Delta E^2 \rangle= (\hbar\omega)^2 \langle( n - \bar n)^2\rangle=(\hbar\omega)^2\sigma^2$; they are directly proportional to   the photon-number variance $\sigma^2=\langle \Delta n^2 \rangle$. Figure 2 presents the mean photon number $\langle n\rangle$ and the  photon autocorrelation function $g^{(2)}(0)$ of the output state, as a function of the scattering ratio $r$. We observe a deterministic increase of the mean photon number (green) above the  thermal value (yellow) for both input mean photon numbers (Fig.~2a), indicating \core{successful  energy extraction}. A maximum energy increase of \el{$11.9(1)\%$} is obtained for \el{$r=14.4(1)\%$} and \el{$\bar n= 3.97(2)$}.  For \el{$r=11.5(1)\%$} and \el{$\bar n= 7.99(3)$}, an increase of \el{$15.8(3)\%$} is achieved. A larger input mean photon number thus leads to a stronger energy enhancement at a smaller scattering ratio. We emphasize that the optimal  values of  the reflectivity \el{(of more than $10\%$ in both cases)} lies outside the usual photon substraction regime $r\ll1$ \cite{nee06,our06,par07,Zavatta2008,Fedorov2015,Bogdanov2017,Hlousek2017,HashemiRafsanjani2017,Katamadze2018}. Meanwhile, positive work extraction  is accompanied by a pronounced suppression of photon correlations (Fig.~2b), below the thermal limit $g^{(2)}(0)_\text{th}=2$ and towards  uncorrelated photons \el{(given by the shot-noise value $g^{(2)}(0)_\text{\el{sn}}=1$)}. This reduction is also amplified for larger input mean photon number $\bar{n}$.
  \begin{figure}[h!]
	\centering
	\includegraphics[width=1.0\linewidth]{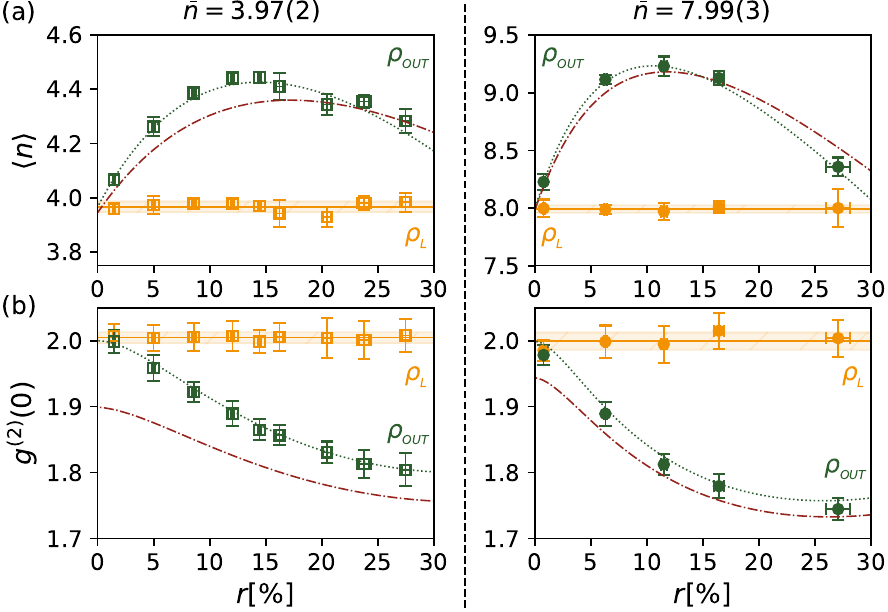}
	\caption{\el{
	(a) Mean photon number $\langle n\rangle$ and (b) second-order correlation function $g^{(2)}(0)$ of the output mode $\rho_\text{out}$ (green) as a function of the scattering ratio $r$ for input thermal states $\rho_\text{L,R}$ (yellow) with mean photon number $\bar{n}=3.97(2)$ (left) and $\bar{n}=7.99(3)$ (right).	Error bars show three standard deviations. 	The measurement-feedforward protocol enhances the mean energy   and reduces   correlations beyond the thermal bound. Quantum statistics boosts $\langle n\rangle$ for moderate $r$ above the corresponding classical limit (red dashed-dotted lines).  
	}}
	\label{fig_2}
\end{figure}

The lowering of the photon correlations  of the output mode has notable consequences as seen in Fig.~3. It indeed increases the photon number fluctuations $\sigma$ \el{for moderate values of $r$} but decreases it  below that of a thermal state \el{for bigger values of $r$} (Fig.~3a). This leads  to a larger mean-to-deviation ratio, $\langle n\rangle/\sigma$, than for the input  mode L, \el{for all reflectivities $r$} (Fig.~3b). We may therefore conclude that the measurement-feedforward protocol enhances the thermodynamic stability of the output state, an important property of small devices subjected to fluctuations \cite{pie18,hol18,den21}. \er{Whereas the output of macroscopic machines is deterministic, and thus stable,  it is stochastic, and hence unstable, for small engines owing to  thermal and quantum fluctuations.} While for thermal states this ratio is given by $\bar{n}/\sigma=1/(1+\bar{n}^{-1})$ \cite{Loudon2010}, and is thus upper bounded by one, we find that this limit is  exceeded for \el{$\bar{n}=7.99(3)$} when \el{$r>6.29(4)\%$}. In this regime, the mean extracted energy  is therefore greater than its  fluctuations.

\cor{In order to highlight the influence of quantum statistics, we next compare   the above results to simulations obtained by replacing the bosonic  
occupation probability $p^\text{L}_n$  of the thermal input mode by its high-temperature (classical) limit, $ p^\text{L}_n = \bar n^n/(\bar n+1)^{n+1} \rightarrow (1/\bar n) \exp(-n/\bar n)$ for $\bar n \gg1$ (Supplemental Material \cite{sup}) (red dashed-dotted lines in Figs.~2 and 3). The  difference between classical and quantum features is clearly visible for both values of $\bar n$ (with a larger difference for the smaller  thermal occupation number $\bar n= 3.97(2)$, as expected). This  demonstrates the nonclassical  nature of the measured data. In particular, quantum statistics enhances the \core{mean} energy output for small and moderate values of the scattering ratio $r$ (Fig.~2a), \core{therefore verifying the predicted boosting effect of bosonic statistics on energy extraction \cite{kim11,cai12,jeo16,ben18,ben18a}}. We further note that classical statistics underestimates the corresponding variance $\sigma$ by not accounting for quantum fluctuations (Fig.~3a) and, as a consequence, overestimates the stability of the device (Fig.~3b).
}
\begin{figure}[t]
	\centering
	\includegraphics[width=1\linewidth]{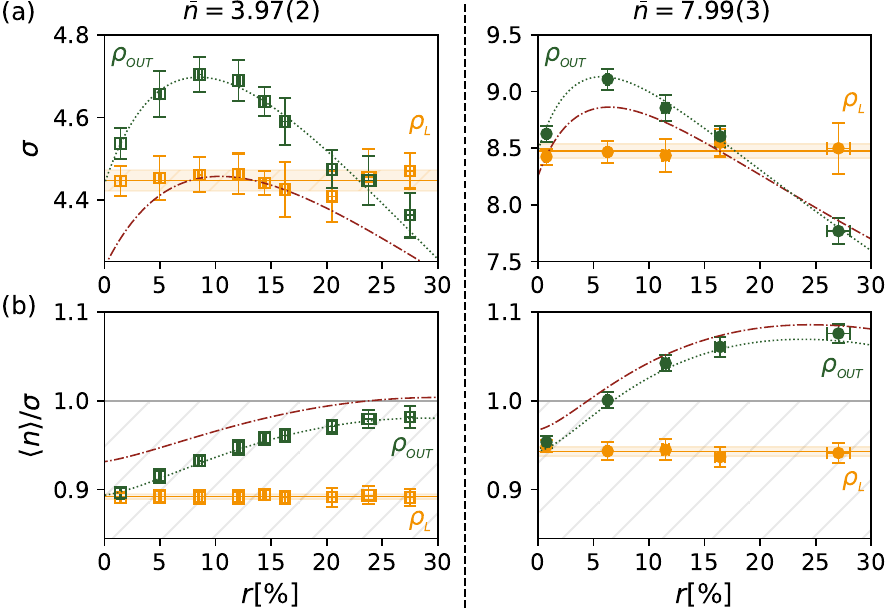}
	\caption{\el{
		(a) Photon number variance $\sigma$ and (b) mean-to-deviation ratio $\langle n\rangle/\sigma$  of the output mode $\rho_\text{out}$ as a function of the scattering ratio $r$ for input thermal states $\rho_\text{L,R}$ with mean photon number $\bar{n}=3.97(2)$ (left) and $\bar{n}=7.99(3)$ (right).
	Same symbols as in Fig.~2.
	The measurement-feedforward  protocol amplifies photon fluctuations $\sigma$ for moderate $r$ and reduces them for larger $r$. The thermodynamic stability of the output mode,  characterized by the mean-to-deviation ratio $\langle n\rangle/\sigma$, is always larger than the corresponding thermal value.	}}
	\label{fig_3}
\end{figure}

\cor{\textit{Nonequilibrium thermodynamic analysis.} Having \core{experimental} access to the photon statistics $p^\text{out}_n$ (Fig.~1b) allows us to perform a detailed thermodynamic analysis of the bosonic  demon. Figure 4a displays the change of entropy, $\Delta S = S(\rho_\text{out}) - S(\rho_\text{L})= -\sum_n (p^\text{out}_n \ln p^\text{out}_n- p^\text{L}_n \ln p^\text{L}_n)$,  induced by the  weak measurement. We observe a behavior that is very similar to that of the mean photon number $\langle n \rangle$ \core{shown in Figs.~2ab}, with a sharp increase for moderate values of $r$ and a decrease for larger scattering ratios. By contrast, the relative entropy \cite{cov06}, $D(\rho_\text{out}||\rho_\text{L})=  \sum_n (p^\text{out}_n \ln p^\text{out}_n- p^\text{out}_n\ln p^\text{L}_n)$,  that quantifies the departure  \core{of the output state from the input thermal state} \cite{def11}, \core{increases linearly for smaller values of $r$ and} saturates for large ones (Fig.~4b). Both information-theoretic   quantities \core{allow us to directly determine} the average energy extracted by the measurement-feedforward protocol via the generalized (reversed) Clausius equality \cite{par89,per02,gav14,may23}
\begin{equation}
\label{2}
\langle \Delta E\rangle = k T_L [\Delta S + D(\rho_\text{out}||\rho_\text{L})],
\end{equation}
 where $k$ is the Boltzmann constant and $T_L$ the temperature of the thermal mode L. Expression \eqref{2}, which is valid arbitrarily far from equilibrium, provides a direct link between the energetic and information-theoretic properties of the quantum demon---and thus insight into his inner operation. In particular, Figs.~4ab reveal that the weak  single-photon measurements  mostly change the entropy  of \core{mode L}, since the contribution of the relative entropy $D(\rho_\text{out}||\rho_\text{L})$ is much smaller than $\Delta S$ (by almost an order of magnitude  for $r\simeq 15\%$). Most of the extracted energy hence stems from the entropic part, and the departure from thermal equilibrium only plays a subdominant role. The first experimental verification of Eq.~\eqref{2} in the quantum regime is presented in Fig.~4\core{c}. Note that, since the relative entropy is nonnegative, Eq.~\eqref{2} implies a reversed Clausius inequality, $\langle \Delta E\rangle \geq k T_L \Delta S$, where  the energy change is larger than the entropy variation times the temperature. The sign of this inequality  differs from that of the usual Clausius inequality which holds for systems  coupled to a  heat bath \cite{par89,per02,gav14,may23}. This observation underscores the unusual quantum thermodynamic features of the many-\core{particle} bosonic demon, \core{that starkly deviate from those of a classical demon}.
}
\begin{figure}[t]
	\centering
	\includegraphics[width=1\linewidth]{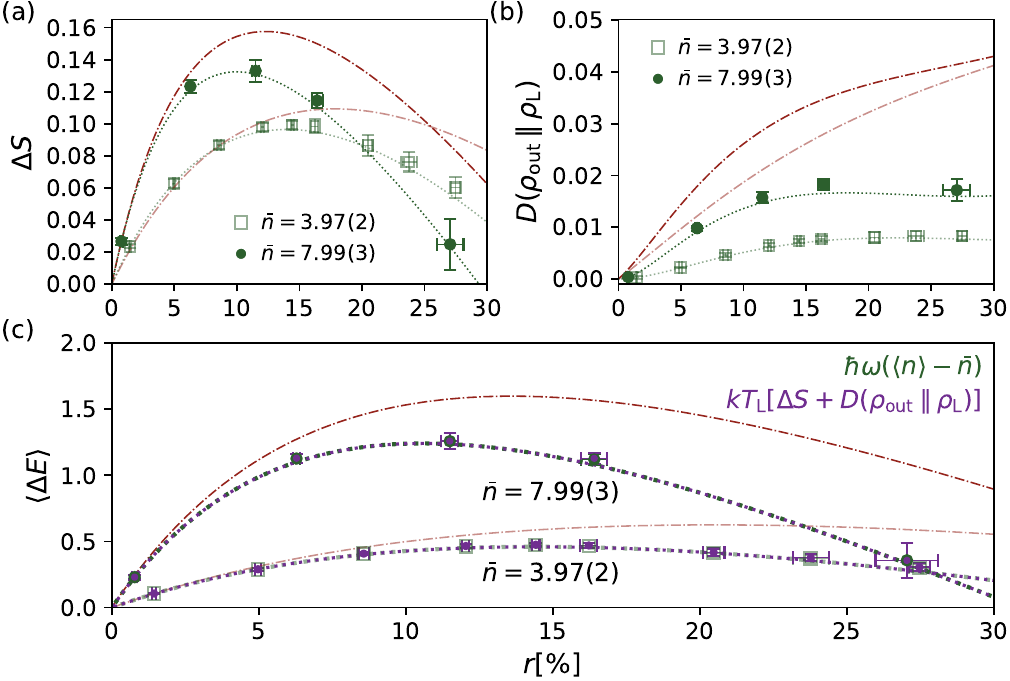}
	\caption{\el{
		(a) Entropy change $\Delta S$ and (b) relative entropy $D(\rho_\text{out}||\rho_\text{L})]$ between input and output modes as a function of the reflectivity $r$ for $\bar{n}=3.94(2)$  and $\bar{n}=7.97(3)$. The entropy change behaves  similarly to the mean photon number $\langle n\rangle$ (Fig.~2a), while the relative entropy saturates.
	Same symbols as in Fig.~2. (c) Confirmation of the generalized (reversed) quantum Clausius equality, Eq.~\eqref{2}, relating thermodynamic and information-theoretic quantities. 	}}
	\label{fig_4}
\end{figure}

\el{\textit{Conclusions.}
\cor{We have  realized a many-\core{particle} photonic  quantum Maxwell demon in a regime \core{dominated by}  bosonic statistics and  discrete photon fluctuations. By combining weak single-photon measurement  and \core{deterministic} feedforward swap operation, we have shown that  the mean \core{extracted} energy and the mean-to-deviation ratio, that is, the stability, can be significantly enhanced and optimized by tuning the reflectivity.  We have additionally  \core{confirmed}  \core{that the many-particle statistics deviates from the classical limit of large mean occupation number}. \core{Remarkably, we have shown that it may enhance the mean energy output above its classical value}. Finally, we have \core{demonstrated} that the mean energy output can be directly determined from information-theoretic quantities, such as entropy and relative entropy, via a generalized (reversed) Clausius equality. This distinguishes our quantum demon from a classical demon. \core{The present methodology could be further extended to different bosonic  systems, like phononic vibrations in quantum optomechanics \cite{Enzian2021} and trapped ions \cite{Um2016}.} Our findings provide unique insight into the nonequilibrium physics of a many-particle Maxwell demon and the general role of quantum statistics as a thermodynamic resource.}}

\begin{acknowledgements}
V.~S. was supported by project CZ.02.01.01/00/22 008/0004649 (QUEENTEC) of EU
and MEYS Czech Republic. 
J.~H. and M.~J. were supported by the Czech Science Foundation under Project 21-18545S, and R.~F. under Project 21-13265X. E.~L. further acknowledges support from the German Science Foundation (DFG) (Project FOR 2724).
\end{acknowledgements}

\onecolumngrid
\appendix
\section{Experimental setup}
\begin{figure*}[h!]
	\centering
	\includegraphics[width=1.0\textwidth]{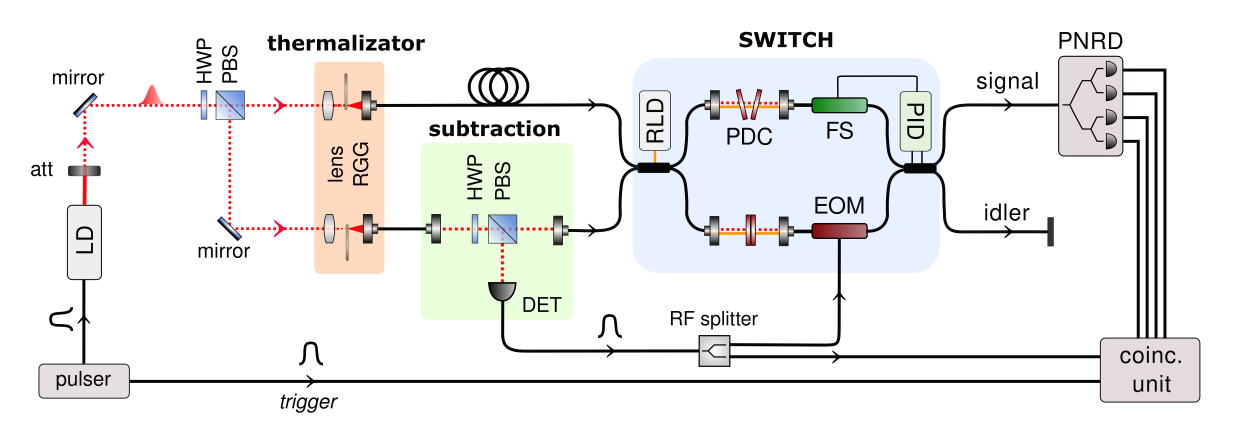}
	\caption{
		Simplified experimental scheme of the bosonic demon.
		Preparation of single-mode thermal states;
		detection of single-photon from the thermal state implemented using a beam splitter with a tunable reflectivity $r$;
		the photonic switch based on Mach-Zehnder interferometer (MZI) with phase dispersion compensator (PDC) and fiber stretcher (FS) used to lock the MZI phase and integrated electro-optic modulator (EOM) controlled by optical feedback for signal switching.
		The output signal is processed by a photon-number-resolving detector (PNRD) and custom coincidence logic (CCU).
	}
	\label{fig_S1}
\end{figure*}

The experimental realization of the quantum photonic Maxwell's demon employs the generation of thermal states of light, photon subtraction, photonic switch, and photon-number-resolving detector.
The scheme of the experimental setup is shown in Fig.~\ref{fig_S1}.
A coherent nanosecond pulsed light is generated by a gain-switched semiconductor laser diode driven by a sub-nanosecond electronic pulse generator with a repetition rate of 2~MHz.
The generated coherent light is split by a 50:50 beam splitter into two coherent states with the same photon number.
The optical intensity of coherent pulses is temporally modulated by two independent rotating ground glasses (RGG) with a random spatial distribution of speckles.
This method is frequently used to generate a quasi-thermal light with virtually perfect Bose-Einstein statistics \cite{Spiller1964}.
A single spatial mode is selected by collecting the scattered light into the single-mode optical fiber.
The result is two identical single-mode thermal states.
One of the thermal states is modified by the subtraction of a single photon.
When the single-photon detection is successful, the photon statistics of the initial thermal state is modified to create an out-of-equilibrium state with a higher mean photon number.
The second thermal state is left unchanged.

These initial states are coupled into the input ports of the 2x2 photonic switch \cite{Svarc2019}.
Fast switching is performed by a low-latency switchable coupler employing a high-visibility fiber Mach-Zehnder interferometer (MZI).
The optical signals can be switched by changing an optical phase using an integrated waveguide electro-optic modulator EOM driven by optical feedback and feedforward circuits using output electronic signal from a single-photon detector.
An active phase stabilization is necessary to keep the random phase fluctuation caused by temperature changes, airflow, and vibrations small enough for advanced long-term measurement.
The real-time phase locking is performed by comparing the output signal to a fixed setpoint and adjusting the phase based on the error signal.
The stabilization technique operating with the strong classical optical reference co-propagating with the initial signal through the MZI was employed to avoid photo-counting noise.

The resulting output signal was analyzed by a photon-number-resolving detector (PNRD).
Our PNRD is based on a spatial-multiplexed scheme.
To measure the multiplexed optical signal, single-photon avalanche diodes (SPAD) were used.
Electronic signals from SPADs are processed by custom coincidence logic \cite{Hlousek2024}.
The probability distribution of the number of photons is evaluated from repeated multi-coincidence measurements.
More details about photon statistics verification can be found in the section below.

\begin{figure}[h!]
	\centering
	\includegraphics[width=1.0\textwidth]{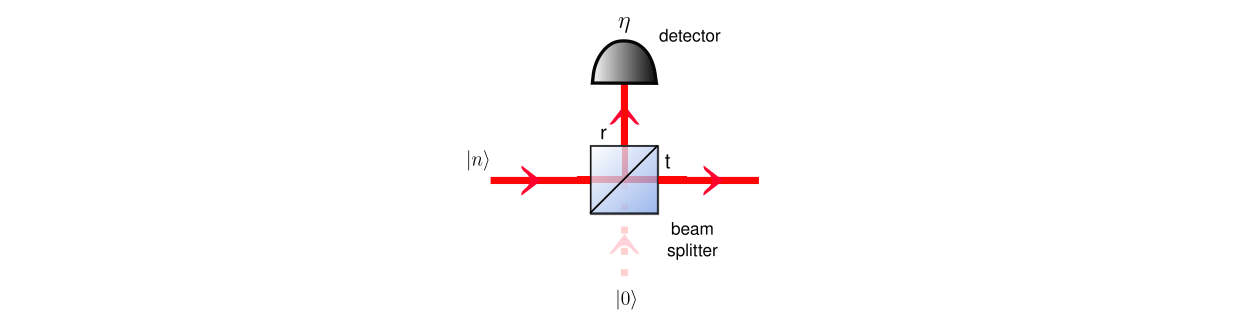}
	\caption{Scheme of the beam splitter and detector setup.}\label{fig:bs_fig}
\end{figure}
\section{SWAP operation}
If $m \geq m_{th}$ photons are detected at least in one $i$-th mode, the demon's measurement is successful, and the SWAP gate flips that mode to the output.
If no photon is detected instead, the demon's measurement is unsuccessful.
In this case, we use the auxiliary mode $A$ with the mean $\bar{n}$ of thermal photons and swap it to the output mode.
A classical SWAP does this feedforward routing for $i$-th mode 
\begin{equation}
F_{i,A}(m) = 
\begin{cases}
\text{SWAP}_{i,out} & \text{if }  m \geq m_{th} \\
\text{SWAP}_{A,out} & \text{if }  m  < m_{th},
\end{cases}
\end{equation} 
where $\text{out}$ denotes the output mode. 
So after applying it, the optical state for $i$-th mode and auxiliary mode $A$ may be written as
\begin{equation}
\rho^{(m_{th})}_{i,A} = \sum_m  F(m)  \left( M(m) S(r) \bar{\rho}_{\text{th}}^{(i)} S^\dagger(r)  M^\dagger(m) \otimes \bar{\rho}_{\text{th}}^{(A)} \right) F^\dagger(m),
\end{equation}
where  $S(r)$ is a unitary beam splitter-like transformation  that characterizes the  scattering process
\begin{equation}
S(r) = e^{-i r \left( \hat{a}^\dagger \hat{b} + \hat{b}^\dagger \hat{a} \right)}
\end{equation}
and $M(m)$ is a positive operator-valued measure (POVM) that describes the  photodetection process
\begin{equation}
    M(m)=\sum_{n=0}^\infty \binom{N_d}{m} \sum_{j=0}^m \binom{m}{j} (-1)^{m-j} \left[(1-\eta)+ \frac{\eta}{N_d}j\right]^n \ket{n}\bra{n}.
\end{equation}

The output state is obtained by taking the respective partial trace over the modes $A$
\begin{equation}
\rho_{\text{out}}^{(m_{th})} = \mbox{Tr}_A \left[\rho^{m_{th}}_{i,A}\right].
\end{equation}

In the experiment, we employed only single-photon subtraction to demonstrate the deterministic preparation of the output state $\rho_{\text{out}}^{(1)}$.

\section{State of the light beam after measurement}
We direct a single light beam towards a beam splitter as shown in Fig.~\ref{fig:bs_fig}. The light beam is then split according to the beam splitter's transmitivity $t$ and reflectivity $r=1-t$, where the reflected light beam is measured via a single photon click counter.  Sperling et al. \cite{Sperling2014} found that in this case of so called photon subtraction, the state of the transmitted light beam after the measurement in Glauber representation is given by
\begin{equation}
	\rho_m = \sum_{l= 0}^\infty \frac{1}{l!} \left[ \binom{N_d}{m} \sum_{j=0}^m \binom{m}{j}
	(-1)^{m-j} \left( - \eta \left(1 - \frac{j}{N_d} \right) \frac{r}{t} \right)^l \right] 
	a^l \Lambda_t^{loss}(\rho_0) 
	 \left(a^{\dagger}\right)^l,
\end{equation}
with $ \Lambda_t^{loss}(\rho_0) = \int d^2 \alpha \frac{1}{t} P_0
 \left( \frac{\alpha}{\sqrt{t}} \right) \ket{\alpha} \bra{\alpha}$ accounting for the losses of the transmitted beam due to the beam splitter. And $P_0(\alpha) = e^{- |\alpha|^2/ \overline{n}} (1/\pi \overline{n})$ describes a thermal input state. Here $\eta$ is the detection efficiency, $N_d$ the total number of detectors, and $\overline{n}$ is initial mean thermal photon number.
After some calculus, we can rewrite the system state in the energy basis. We obtain  after photon subtraction,

\begin{equation}\label{eq:rho_after}
	 \rho_m' = \binom{N_d}{m} \sum_{n=0}^\infty \sum^m_{j=0} \binom{m}{j} (-1)^{m-j} \frac{1}{\overline{n} t}
	  \left( \frac{1}{1+ 1/\overline{n} t  +  \eta r/t (1 - \frac{j}{N_d})  }  \right)^{n+1}  \ket{n} \bra{n}.
\end{equation}
Note that $\rho_m'$ is not normalized yet, but we still have to consider the probability $p_m$ of
detecting $m$ photons :
\begin{equation}\label{eq:p(m)}
	p(m) = Tr \left[ \rho'_m  \right] = \binom{N_d}{m} \sum^m_{j=0} \binom{m}{j} (-1)^{m-j} \frac{1}{1+ \eta r \overline{n}(1 - \frac{j}{N_d})} .
\end{equation}
From Eq. \ref{eq:rho_after}, the new probability distribution of the main system can be easily seen to be
\begin{equation}\label{eq:pnm_after}
	p(n|m) = \frac{1}{p(m)} \binom{N_d}{m} \sum_{j=0}^m \binom{m}{j} (-1)^{m-j} \frac{1}{\overline{n} t} \frac{1}{\lambda_j^{n+1}},
\end{equation}
with $\lambda_j =1+ 1/\overline{n} t  +  \eta r/t (1 - \frac{j}{N_d})  $.\\

\section{High-temperature (classical) limit}
In Figs.~2 and 3 of the main text, we compare our approach with the high-temperature (classical) limit of the light beams, while maintaining the same average photon number $\bar{n}$. Below, we present the resulting probability distributions for this high-temperature limit, illustrating how our quantum approach contrasts with the classical regime.

The input thermal distribution in this limit is defined as: 
\begin{equation}
p_{\text{th}}(n) = \frac{\bar{n}^{n}}{(\bar{n}+1)^{n+1}} \approx \frac{1}{\bar{n}} e^{-n/\bar{n}}. \tag{S1}
\end{equation}

Using this limit, we obtain:
\begin{enumerate}
    \item[(i)] the post-photon subtraction state
    \begin{equation}
    \rho_{\text{high}, m}' = \binom{N_d}{m} \sum_{n=0}^{\infty} \sum_{j=0}^{m} \binom{m}{j} (-1)^{m-j} \frac{1}{\bar{n}t} \left( \frac{1}{1 + \eta r / t (1 - j / N_d)} \right)^{n+1} |n\rangle \langle n| . \tag{S2}
    \end{equation}

    \item[(ii)] the probability $p_{\text{high}}(m)$ of detecting $m$ photons
    \begin{equation}
    p_{\text{high}}(m) = \text{Tr} [\rho_m'] = \binom{N_d}{m} \sum_{j=0}^{m} \binom{m}{j} (-1)^{m-j} \frac{1}{\eta r \bar{n} (1 - j / N_d)}. \tag{S3}
    \end{equation}

    \item[(iii)] the new probability distribution of the main system
    \begin{equation}
    p_{\text{high}}(n|m) = \frac{1}{p(m)} \binom{N_d}{m} \sum_{j=0}^{m} \binom{m}{j} (-1)^{m-j} \frac{1}{\bar{n}t} \frac{1}{\lambda_j^{n+1}}, \tag{S4}
    \end{equation}
\end{enumerate}

with $\lambda_j = 1 + \eta r / t (1 - j / N_d)$.

\begin{figure*}[ht]
	\centering
	\includegraphics[width=1.0\textwidth]{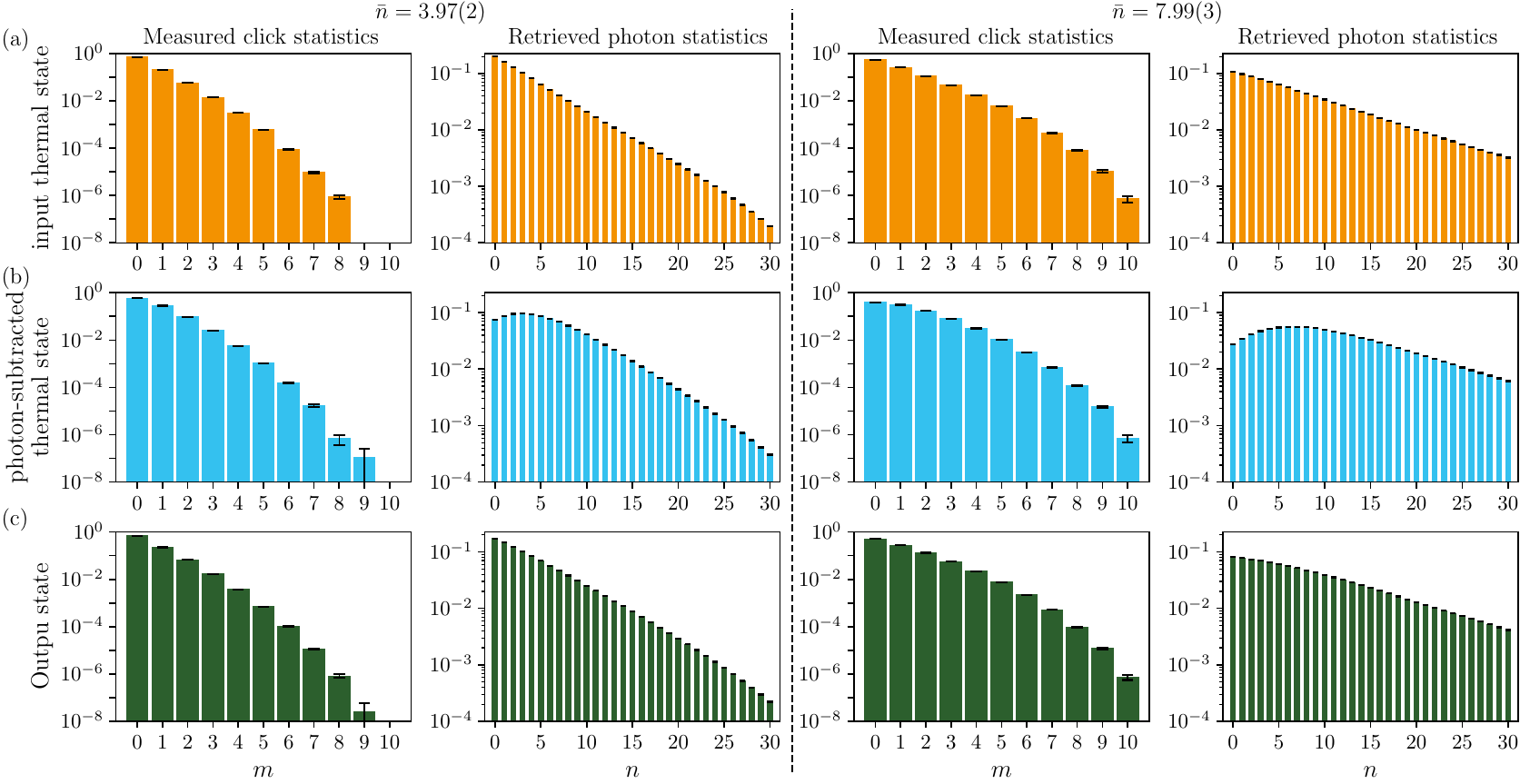}
	\caption{Measured click statistics and retrieved photon statistics of (a) the input single-mode thermal state, (b) one-photon-subtracted single-mode thermal state, and (c) deterministic prepared output state for mean photon numbers: $\bar{n}$ = 3.97(2) $r$ = 14.4(1)\% (left) and $\bar{n}$ = 7.99(3) and $r$ = 11.5(1)\% (right).
    }
	\label{fig_S2}
\end{figure*}
\section{Photon statistics retrieval}
The output mode is divided into an optical network consisting of cascaded tunable beam splitters and processed by single-photon avalanche diodes. Subsequent data processing is based on a real-time measurement of all the coincidence events between the detection channels. Due to non-unity detection efficiency, noise, and a finite number of single-photon detection channels, we observe the probability distribution of the coincidence events (i.e. click statistics) $c_m$ instead of the photon statistics $p_n$. To estimate the photon statistics of detected light, the expectation-maximization-entropy algorithm using entropy regularization of the popular maximum-likelihood expectation-maximization technique is implemented. A detailed derivation and the Python code are provided in the supplementary material of \cite{Hlousek2019} and accessible via GitHub \cite{github}. This detection scheme and photon statistics retrieval are both universal and suitable for all types/kinds of initial photon states.

Figure~\ref{fig_S2} illustrates measured click statistics and retrieved photon statistics of the input single-mode thermal state, one-photon-subtracted single-mode thermal state, and deterministic prepared output state. The results are shown for mean photon numbers $\bar{n} = 3.97(2)$ and $r = 14.4(1)\%$ and $\bar{n} = 7.99(3)$ and $r = 11.5(1)\%$, corresponding to the left and right panels, respectively. We analyzed photon statistics retrieval accuracy, evaluating the fidelities of all initial states, achieving $\mathcal{F} > 0.998$.

\twocolumngrid
\bibliography{bibliography}

\begin{thebibliography}{67}%
\makeatletter
\providecommand \@ifxundefined [1]{%
 \@ifx{#1\undefined}
}%
\providecommand \@ifnum [1]{%
 \ifnum #1\expandafter \@firstoftwo
 \else \expandafter \@secondoftwo
 \fi
}%
\providecommand \@ifx [1]{%
 \ifx #1\expandafter \@firstoftwo
 \else \expandafter \@secondoftwo
 \fi
}%
\providecommand \natexlab [1]{#1}%
\providecommand \enquote  [1]{``#1''}%
\providecommand \bibnamefont  [1]{#1}%
\providecommand \bibfnamefont [1]{#1}%
\providecommand \citenamefont [1]{#1}%
\providecommand \href@noop [0]{\@secondoftwo}%
\providecommand \href [0]{\begingroup \@sanitize@url \@href}%
\providecommand \@href[1]{\@@startlink{#1}\@@href}%
\providecommand \@@href[1]{\endgroup#1\@@endlink}%
\providecommand \@sanitize@url [0]{\catcode `\\12\catcode `\$12\catcode `\&12\catcode `\#12\catcode `\^12\catcode `\_12\catcode `\%12\relax}%
\providecommand \@@startlink[1]{}%
\providecommand \@@endlink[0]{}%
\providecommand \url  [0]{\begingroup\@sanitize@url \@url }%
\providecommand \@url [1]{\endgroup\@href {#1}{\urlprefix }}%
\providecommand \urlprefix  [0]{URL }%
\providecommand \Eprint [0]{\href }%
\providecommand \doibase [0]{https://doi.org/}%
\providecommand \selectlanguage [0]{\@gobble}%
\providecommand \bibinfo  [0]{\@secondoftwo}%
\providecommand \bibfield  [0]{\@secondoftwo}%
\providecommand \translation [1]{[#1]}%
\providecommand \BibitemOpen [0]{}%
\providecommand \bibitemStop [0]{}%
\providecommand \bibitemNoStop [0]{.\EOS\space}%
\providecommand \EOS [0]{\spacefactor3000\relax}%
\providecommand \BibitemShut  [1]{\csname bibitem#1\endcsname}%
\let\auto@bib@innerbib\@empty
\bibitem [{\citenamefont {Leff}\ and\ \citenamefont {Rex}(2003)}]{lef03}%
  \BibitemOpen
  \bibfield  {author} {\bibinfo {author} {\bibfnamefont {H.~S.}\ \bibnamefont {Leff}}\ and\ \bibinfo {author} {\bibfnamefont {A.~F.}\ \bibnamefont {Rex}},\ }\href@noop {} {\emph {\bibinfo {title} {Maxwell's Demon 2: Entropy, Classical and Quantum Information, Computing}}}\ (\bibinfo  {publisher} {Institute of Physics Publishing},\ \bibinfo {year} {2003})\BibitemShut {NoStop}%
\bibitem [{\citenamefont {Plenio}\ and\ \citenamefont {Vitelli}(2001)}]{ple01}%
  \BibitemOpen
  \bibfield  {author} {\bibinfo {author} {\bibfnamefont {M.~B.}\ \bibnamefont {Plenio}}\ and\ \bibinfo {author} {\bibfnamefont {V.}~\bibnamefont {Vitelli}},\ }\bibfield  {title} {\bibinfo {title} {The physics of forgetting: Landauer's erasure principle and information theory},\ }\href@noop {} {\bibfield  {journal} {\bibinfo  {journal} {Contemporary Phys.}\ }\textbf {\bibinfo {volume} {42}},\ \bibinfo {pages} {25} (\bibinfo {year} {2001})}\BibitemShut {NoStop}%
\bibitem [{\citenamefont {Maruyama}\ \emph {et~al.}(2009)\citenamefont {Maruyama}, \citenamefont {Nori},\ and\ \citenamefont {Vedral}}]{mar09}%
  \BibitemOpen
  \bibfield  {author} {\bibinfo {author} {\bibfnamefont {K.}~\bibnamefont {Maruyama}}, \bibinfo {author} {\bibfnamefont {F.}~\bibnamefont {Nori}},\ and\ \bibinfo {author} {\bibfnamefont {V.}~\bibnamefont {Vedral}},\ }\bibfield  {title} {\bibinfo {title} {Colloquium: The physics of {M}axwell's demon and information},\ }\href@noop {} {\bibfield  {journal} {\bibinfo  {journal} {Rev. Mod. Phys.}\ }\textbf {\bibinfo {volume} {81}},\ \bibinfo {pages} {1} (\bibinfo {year} {2009})}\BibitemShut {NoStop}%
\bibitem [{\citenamefont {Parrondo}\ \emph {et~al.}(2015)\citenamefont {Parrondo}, \citenamefont {Horowitz},\ and\ \citenamefont {Sagawa}}]{par15}%
  \BibitemOpen
  \bibfield  {author} {\bibinfo {author} {\bibfnamefont {J.}~\bibnamefont {Parrondo}}, \bibinfo {author} {\bibfnamefont {J.}~\bibnamefont {Horowitz}},\ and\ \bibinfo {author} {\bibfnamefont {T.}~\bibnamefont {Sagawa}},\ }\bibfield  {title} {\bibinfo {title} {Thermodynamics of information},\ }\href@noop {} {\bibfield  {journal} {\bibinfo  {journal} {Nature Phys.}\ }\textbf {\bibinfo {volume} {11}},\ \bibinfo {pages} {131} (\bibinfo {year} {2015})}\BibitemShut {NoStop}%
\bibitem [{\citenamefont {Lutz}\ and\ \citenamefont {Ciliberto}(2015)}]{lut15}%
  \BibitemOpen
  \bibfield  {author} {\bibinfo {author} {\bibfnamefont {E.}~\bibnamefont {Lutz}}\ and\ \bibinfo {author} {\bibfnamefont {S.}~\bibnamefont {Ciliberto}},\ }\bibfield  {title} {\bibinfo {title} {Information: From {M}axwell's demon to {L}andauer's eraser},\ }\href@noop {} {\bibfield  {journal} {\bibinfo  {journal} {Physics Today}\ }\textbf {\bibinfo {volume} {68}},\ \bibinfo {pages} {30} (\bibinfo {year} {2015})}\BibitemShut {NoStop}%
\bibitem [{\citenamefont {Raizen}(2009)}]{rai09}%
  \BibitemOpen
  \bibfield  {author} {\bibinfo {author} {\bibfnamefont {M.~G.}\ \bibnamefont {Raizen}},\ }\bibfield  {title} {\bibinfo {title} {Comprehensive control of atomic motion},\ }\href@noop {} {\bibfield  {journal} {\bibinfo  {journal} {Science}\ }\textbf {\bibinfo {volume} {324}},\ \bibinfo {pages} {1403} (\bibinfo {year} {2009})}\BibitemShut {NoStop}%
\bibitem [{\citenamefont {Toyabe}\ \emph {et~al.}(2010)\citenamefont {Toyabe}, \citenamefont {Sagawa}, \citenamefont {Ueda}, \citenamefont {Muneyuki},\ and\ \citenamefont {Sano}}]{toy10}%
  \BibitemOpen
  \bibfield  {author} {\bibinfo {author} {\bibfnamefont {S.}~\bibnamefont {Toyabe}}, \bibinfo {author} {\bibfnamefont {T.}~\bibnamefont {Sagawa}}, \bibinfo {author} {\bibfnamefont {M.}~\bibnamefont {Ueda}}, \bibinfo {author} {\bibfnamefont {E.}~\bibnamefont {Muneyuki}},\ and\ \bibinfo {author} {\bibfnamefont {M.}~\bibnamefont {Sano}},\ }\bibfield  {title} {\bibinfo {title} {Experimental demonstration of information-to-energy conversion and validation of the generalized {J}arzynski equality},\ }\href@noop {} {\bibfield  {journal} {\bibinfo  {journal} {Nature Phys.}\ }\textbf {\bibinfo {volume} {6}},\ \bibinfo {pages} {988} (\bibinfo {year} {2010})}\BibitemShut {NoStop}%
\bibitem [{\citenamefont {Roldan}\ \emph {et~al.}(2014)\citenamefont {Roldan}, \citenamefont {Martinez}, \citenamefont {Parrondo},\ and\ \citenamefont {Petrov}}]{rol14}%
  \BibitemOpen
  \bibfield  {author} {\bibinfo {author} {\bibfnamefont {E.}~\bibnamefont {Roldan}}, \bibinfo {author} {\bibfnamefont {I.~A.}\ \bibnamefont {Martinez}}, \bibinfo {author} {\bibfnamefont {J.~M.~R.}\ \bibnamefont {Parrondo}},\ and\ \bibinfo {author} {\bibfnamefont {D.}~\bibnamefont {Petrov}},\ }\bibfield  {title} {\bibinfo {title} {Universal features in the energetics of symmetry breaking},\ }\href@noop {} {\bibfield  {journal} {\bibinfo  {journal} {Nature Phys.}\ }\textbf {\bibinfo {volume} {10}},\ \bibinfo {pages} {457} (\bibinfo {year} {2014})}\BibitemShut {NoStop}%
\bibitem [{\citenamefont {Koski}\ \emph {et~al.}(2014{\natexlab{a}})\citenamefont {Koski}, \citenamefont {Maisi}, \citenamefont {Pekola},\ and\ \citenamefont {Averin}}]{kos14}%
  \BibitemOpen
  \bibfield  {author} {\bibinfo {author} {\bibfnamefont {J.~V.}\ \bibnamefont {Koski}}, \bibinfo {author} {\bibfnamefont {V.~F.}\ \bibnamefont {Maisi}}, \bibinfo {author} {\bibfnamefont {J.~P.}\ \bibnamefont {Pekola}},\ and\ \bibinfo {author} {\bibfnamefont {D.~V.}\ \bibnamefont {Averin}},\ }\bibfield  {title} {\bibinfo {title} {Experimental realization of a {S}zilard engine with a single electron},\ }\href@noop {} {\bibfield  {journal} {\bibinfo  {journal} {Proc. Natl. Acad. Sci. U.S.A}\ }\textbf {\bibinfo {volume} {111}},\ \bibinfo {pages} {13786} (\bibinfo {year} {2014}{\natexlab{a}})}\BibitemShut {NoStop}%
\bibitem [{\citenamefont {Koski}\ \emph {et~al.}(2014{\natexlab{b}})\citenamefont {Koski}, \citenamefont {Maisi}, \citenamefont {Sagawa},\ and\ \citenamefont {Pekola}}]{kos14a}%
  \BibitemOpen
  \bibfield  {author} {\bibinfo {author} {\bibfnamefont {J.~V.}\ \bibnamefont {Koski}}, \bibinfo {author} {\bibfnamefont {V.~F.}\ \bibnamefont {Maisi}}, \bibinfo {author} {\bibfnamefont {T.}~\bibnamefont {Sagawa}},\ and\ \bibinfo {author} {\bibfnamefont {J.~P.}\ \bibnamefont {Pekola}},\ }\bibfield  {title} {\bibinfo {title} {Experimental observation of the role of mutual information in the nonequilibrium dynamics of a {M}axwell demon},\ }\href@noop {} {\bibfield  {journal} {\bibinfo  {journal} {Phys. Rev. Lett.}\ }\textbf {\bibinfo {volume} {113}},\ \bibinfo {pages} {030601} (\bibinfo {year} {2014}{\natexlab{b}})}\BibitemShut {NoStop}%
\bibitem [{\citenamefont {Koski}\ \emph {et~al.}(2015)\citenamefont {Koski}, \citenamefont {Kutvonen}, \citenamefont {Khaymovich}, \citenamefont {Ala-Nissila},\ and\ \citenamefont {Pekola}}]{kos15}%
  \BibitemOpen
  \bibfield  {author} {\bibinfo {author} {\bibfnamefont {J.~V.}\ \bibnamefont {Koski}}, \bibinfo {author} {\bibfnamefont {A.}~\bibnamefont {Kutvonen}}, \bibinfo {author} {\bibfnamefont {I.~M.}\ \bibnamefont {Khaymovich}}, \bibinfo {author} {\bibfnamefont {T.}~\bibnamefont {Ala-Nissila}},\ and\ \bibinfo {author} {\bibfnamefont {J.~P.}\ \bibnamefont {Pekola}},\ }\bibfield  {title} {\bibinfo {title} {On-chip {M}axwell's demon as an information-powered refrigerator},\ }\href@noop {} {\bibfield  {journal} {\bibinfo  {journal} {Phys. Rev. Lett.}\ }\textbf {\bibinfo {volume} {115}},\ \bibinfo {pages} {260602} (\bibinfo {year} {2015})}\BibitemShut {NoStop}%
\bibitem [{\citenamefont {Chida}\ \emph {et~al.}(2017)\citenamefont {Chida}, \citenamefont {Desai}, \citenamefont {Nishiguchi},\ and\ \citenamefont {Fujiwara}}]{chi17}%
  \BibitemOpen
  \bibfield  {author} {\bibinfo {author} {\bibfnamefont {K.}~\bibnamefont {Chida}}, \bibinfo {author} {\bibfnamefont {S.}~\bibnamefont {Desai}}, \bibinfo {author} {\bibfnamefont {K.}~\bibnamefont {Nishiguchi}},\ and\ \bibinfo {author} {\bibfnamefont {A.}~\bibnamefont {Fujiwara}},\ }\bibfield  {title} {\bibinfo {title} {Power generator driven by {M}axwell's demon},\ }\href@noop {} {\bibfield  {journal} {\bibinfo  {journal} {Nature Comm.}\ }\textbf {\bibinfo {volume} {8}},\ \bibinfo {pages} {15310} (\bibinfo {year} {2017})}\BibitemShut {NoStop}%
\bibitem [{\citenamefont {Kumar}\ \emph {et~al.}(2018)\citenamefont {Kumar}, \citenamefont {Wu}, \citenamefont {Giraldo},\ and\ \citenamefont {Weiss}}]{kum18}%
  \BibitemOpen
  \bibfield  {author} {\bibinfo {author} {\bibfnamefont {A.}~\bibnamefont {Kumar}}, \bibinfo {author} {\bibfnamefont {T.~Y.}\ \bibnamefont {Wu}}, \bibinfo {author} {\bibfnamefont {F.}~\bibnamefont {Giraldo}},\ and\ \bibinfo {author} {\bibfnamefont {D.~S.}\ \bibnamefont {Weiss}},\ }\bibfield  {title} {\bibinfo {title} {Sorting ultracold atoms in a three-dimensional optical lattice in a realization of {M}axwell's demon},\ }\href@noop {} {\bibfield  {journal} {\bibinfo  {journal} {Nature}\ }\textbf {\bibinfo {volume} {561}},\ \bibinfo {pages} {83} (\bibinfo {year} {2018})}\BibitemShut {NoStop}%
\bibitem [{\citenamefont {Paneru}\ \emph {et~al.}(2018)\citenamefont {Paneru}, \citenamefont {Lee}, \citenamefont {Tlusty},\ and\ \citenamefont {Pak}}]{pan18}%
  \BibitemOpen
  \bibfield  {author} {\bibinfo {author} {\bibfnamefont {G.}~\bibnamefont {Paneru}}, \bibinfo {author} {\bibfnamefont {D.~Y.}\ \bibnamefont {Lee}}, \bibinfo {author} {\bibfnamefont {T.}~\bibnamefont {Tlusty}},\ and\ \bibinfo {author} {\bibfnamefont {H.~K.}\ \bibnamefont {Pak}},\ }\bibfield  {title} {\bibinfo {title} {Lossless {B}rownian information engine},\ }\href@noop {} {\bibfield  {journal} {\bibinfo  {journal} {Phys. Rev. Lett.}\ }\textbf {\bibinfo {volume} {120}},\ \bibinfo {pages} {020601} (\bibinfo {year} {2018})}\BibitemShut {NoStop}%
\bibitem [{\citenamefont {Admon}\ \emph {et~al.}(2018)\citenamefont {Admon}, \citenamefont {Rahav},\ and\ \citenamefont {Roichman}}]{adm18}%
  \BibitemOpen
  \bibfield  {author} {\bibinfo {author} {\bibfnamefont {T.}~\bibnamefont {Admon}}, \bibinfo {author} {\bibfnamefont {S.}~\bibnamefont {Rahav}},\ and\ \bibinfo {author} {\bibfnamefont {Y.}~\bibnamefont {Roichman}},\ }\bibfield  {title} {\bibinfo {title} {Experimental realization of an information machine with tunable temporal correlations},\ }\href@noop {} {\bibfield  {journal} {\bibinfo  {journal} {Phys. Rev. Lett.}\ }\textbf {\bibinfo {volume} {121}},\ \bibinfo {pages} {180601} (\bibinfo {year} {2018})}\BibitemShut {NoStop}%
\bibitem [{\citenamefont {Ribezzi-Crivellari}\ and\ \citenamefont {Ritort}(2019)}]{rib19}%
  \BibitemOpen
  \bibfield  {author} {\bibinfo {author} {\bibfnamefont {M.}~\bibnamefont {Ribezzi-Crivellari}}\ and\ \bibinfo {author} {\bibfnamefont {F.}~\bibnamefont {Ritort}},\ }\bibfield  {title} {\bibinfo {title} {Large work extraction and the {L}andauer limit in a continuous {M}axwell demon},\ }\href@noop {} {\bibfield  {journal} {\bibinfo  {journal} {Nature Phys.}\ }\textbf {\bibinfo {volume} {15}},\ \bibinfo {pages} {660} (\bibinfo {year} {2019})}\BibitemShut {NoStop}%
\bibitem [{\citenamefont {Debiossac}\ \emph {et~al.}(2020)\citenamefont {Debiossac}, \citenamefont {Grass}, \citenamefont {Alonso}, \citenamefont {Lutz},\ and\ \citenamefont {Kiesel}}]{deb20}%
  \BibitemOpen
  \bibfield  {author} {\bibinfo {author} {\bibfnamefont {M.}~\bibnamefont {Debiossac}}, \bibinfo {author} {\bibfnamefont {D.}~\bibnamefont {Grass}}, \bibinfo {author} {\bibfnamefont {J.~J.}\ \bibnamefont {Alonso}}, \bibinfo {author} {\bibfnamefont {E.}~\bibnamefont {Lutz}},\ and\ \bibinfo {author} {\bibfnamefont {N.}~\bibnamefont {Kiesel}},\ }\bibfield  {title} {\bibinfo {title} {Thermodynamics of continuous non-{M}arkovian feedback control},\ }\href@noop {} {\bibfield  {journal} {\bibinfo  {journal} {Nature Comm.}\ }\textbf {\bibinfo {volume} {11}},\ \bibinfo {pages} {1360} (\bibinfo {year} {2020})}\BibitemShut {NoStop}%
\bibitem [{\citenamefont {Saha}\ \emph {et~al.}(2021)\citenamefont {Saha}, \citenamefont {Lucero}, \citenamefont {Ehrich}, \citenamefont {Sivak},\ and\ \citenamefont {Bechhoefer}}]{sah21}%
  \BibitemOpen
  \bibfield  {author} {\bibinfo {author} {\bibfnamefont {T.~K.}\ \bibnamefont {Saha}}, \bibinfo {author} {\bibfnamefont {J.~N.~E.}\ \bibnamefont {Lucero}}, \bibinfo {author} {\bibfnamefont {J.}~\bibnamefont {Ehrich}}, \bibinfo {author} {\bibfnamefont {D.~A.}\ \bibnamefont {Sivak}},\ and\ \bibinfo {author} {\bibfnamefont {J.}~\bibnamefont {Bechhoefer}},\ }\bibfield  {title} {\bibinfo {title} {Maximizing power and velocity of an information engine},\ }\href@noop {} {\bibfield  {journal} {\bibinfo  {journal} {Proc. Natl. Acad. Sci. U.S.A.}\ }\textbf {\bibinfo {volume} {118}},\ \bibinfo {pages} {e2023356118} (\bibinfo {year} {2021})}\BibitemShut {NoStop}%
\bibitem [{\citenamefont {Saha}\ \emph {et~al.}(2023)\citenamefont {Saha}, \citenamefont {Ehrich}, \citenamefont {Gavrilov}, \citenamefont {Still}, \citenamefont {Sivak},\ and\ \citenamefont {Bechhoefer}}]{sah23}%
  \BibitemOpen
  \bibfield  {author} {\bibinfo {author} {\bibfnamefont {T.~K.}\ \bibnamefont {Saha}}, \bibinfo {author} {\bibfnamefont {J.}~\bibnamefont {Ehrich}}, \bibinfo {author} {\bibfnamefont {M.~c.~v.}\ \bibnamefont {Gavrilov}}, \bibinfo {author} {\bibfnamefont {S.}~\bibnamefont {Still}}, \bibinfo {author} {\bibfnamefont {D.~A.}\ \bibnamefont {Sivak}},\ and\ \bibinfo {author} {\bibfnamefont {J.}~\bibnamefont {Bechhoefer}},\ }\bibfield  {title} {\bibinfo {title} {Information engine in a nonequilibrium bath},\ }\href@noop {} {\bibfield  {journal} {\bibinfo  {journal} {Phys. Rev. Lett.}\ }\textbf {\bibinfo {volume} {131}},\ \bibinfo {pages} {057101} (\bibinfo {year} {2023})}\BibitemShut {NoStop}%
\bibitem [{\citenamefont {Yan}\ \emph {et~al.}(2024)\citenamefont {Yan}, \citenamefont {Bu}, \citenamefont {Zeng}, \citenamefont {Zhang}, \citenamefont {Cui}, \citenamefont {Zhou}, \citenamefont {Su}, \citenamefont {Chen}, \citenamefont {Wang}, \citenamefont {Chen},\ and\ \citenamefont {Feng}}]{yan24}%
  \BibitemOpen
  \bibfield  {author} {\bibinfo {author} {\bibfnamefont {L.-L.}\ \bibnamefont {Yan}}, \bibinfo {author} {\bibfnamefont {J.-T.}\ \bibnamefont {Bu}}, \bibinfo {author} {\bibfnamefont {Q.}~\bibnamefont {Zeng}}, \bibinfo {author} {\bibfnamefont {K.}~\bibnamefont {Zhang}}, \bibinfo {author} {\bibfnamefont {K.-F.}\ \bibnamefont {Cui}}, \bibinfo {author} {\bibfnamefont {F.}~\bibnamefont {Zhou}}, \bibinfo {author} {\bibfnamefont {S.-L.}\ \bibnamefont {Su}}, \bibinfo {author} {\bibfnamefont {L.}~\bibnamefont {Chen}}, \bibinfo {author} {\bibfnamefont {J.}~\bibnamefont {Wang}}, \bibinfo {author} {\bibfnamefont {G.}~\bibnamefont {Chen}},\ and\ \bibinfo {author} {\bibfnamefont {M.}~\bibnamefont {Feng}},\ }\bibfield  {title} {\bibinfo {title} {Experimental verification of demon-involved fluctuation theorems},\ }\href@noop {} {\bibfield  {journal} {\bibinfo  {journal} {Phys. Rev. Lett.}\ }\textbf {\bibinfo {volume} {133}},\ \bibinfo {pages} {090402} (\bibinfo {year} {2024})}\BibitemShut {NoStop}%
\bibitem [{\citenamefont {Camati}\ \emph {et~al.}(2016)\citenamefont {Camati}, \citenamefont {Peterson}, \citenamefont {Batalh\~ao}, \citenamefont {Micadei}, \citenamefont {Souza}, \citenamefont {Sarthour}, \citenamefont {Oliveira},\ and\ \citenamefont {Serra}}]{cam16}%
  \BibitemOpen
  \bibfield  {author} {\bibinfo {author} {\bibfnamefont {P.~A.}\ \bibnamefont {Camati}}, \bibinfo {author} {\bibfnamefont {J.~P.~S.}\ \bibnamefont {Peterson}}, \bibinfo {author} {\bibfnamefont {T.~B.}\ \bibnamefont {Batalh\~ao}}, \bibinfo {author} {\bibfnamefont {K.}~\bibnamefont {Micadei}}, \bibinfo {author} {\bibfnamefont {A.~M.}\ \bibnamefont {Souza}}, \bibinfo {author} {\bibfnamefont {R.~S.}\ \bibnamefont {Sarthour}}, \bibinfo {author} {\bibfnamefont {I.~S.}\ \bibnamefont {Oliveira}},\ and\ \bibinfo {author} {\bibfnamefont {R.~M.}\ \bibnamefont {Serra}},\ }\bibfield  {title} {\bibinfo {title} {Experimental rectification of entropy production by {M}axwell's demon in a quantum system},\ }\href@noop {} {\bibfield  {journal} {\bibinfo  {journal} {Phys. Rev. Lett.}\ }\textbf {\bibinfo {volume} {117}},\ \bibinfo {pages} {240502} (\bibinfo {year} {2016})}\BibitemShut {NoStop}%
\bibitem [{\citenamefont {Cottet}\ \emph {et~al.}(2017)\citenamefont {Cottet}, \citenamefont {Jezouin}, \citenamefont {Bretheau}, \citenamefont {Campagne-Ibarcq}, \citenamefont {Ficheux}, \citenamefont {Anders}, \citenamefont {Auff\`eves}, \citenamefont {Azouit}, \citenamefont {Rouchon},\ and\ \citenamefont {Huard}}]{cot17}%
  \BibitemOpen
  \bibfield  {author} {\bibinfo {author} {\bibfnamefont {N.}~\bibnamefont {Cottet}}, \bibinfo {author} {\bibfnamefont {S.}~\bibnamefont {Jezouin}}, \bibinfo {author} {\bibfnamefont {L.}~\bibnamefont {Bretheau}}, \bibinfo {author} {\bibfnamefont {P.}~\bibnamefont {Campagne-Ibarcq}}, \bibinfo {author} {\bibfnamefont {Q.}~\bibnamefont {Ficheux}}, \bibinfo {author} {\bibfnamefont {J.}~\bibnamefont {Anders}}, \bibinfo {author} {\bibfnamefont {A.}~\bibnamefont {Auff\`eves}}, \bibinfo {author} {\bibfnamefont {R.}~\bibnamefont {Azouit}}, \bibinfo {author} {\bibfnamefont {P.}~\bibnamefont {Rouchon}},\ and\ \bibinfo {author} {\bibfnamefont {B.}~\bibnamefont {Huard}},\ }\bibfield  {title} {\bibinfo {title} {Observing a quantum {M}axwell demon at work},\ }\href@noop {} {\bibfield  {journal} {\bibinfo  {journal} {Proc. Natl. Acad. Sci. U.S.A.}\ }\textbf {\bibinfo {volume} {114}},\ \bibinfo {pages} {7561} (\bibinfo {year} {2017})}\BibitemShut {NoStop}%
\bibitem [{\citenamefont {Masuyama}\ \emph {et~al.}(2018)\citenamefont {Masuyama}, \citenamefont {Funo}, \citenamefont {Murashita}, \citenamefont {Noguchi}, \citenamefont {Kono}, \citenamefont {Tabuchi}, \citenamefont {Yamazaki}, \citenamefont {Ueda},\ and\ \citenamefont {Nakamura}}]{mas18}%
  \BibitemOpen
  \bibfield  {author} {\bibinfo {author} {\bibfnamefont {Y.}~\bibnamefont {Masuyama}}, \bibinfo {author} {\bibfnamefont {K.}~\bibnamefont {Funo}}, \bibinfo {author} {\bibfnamefont {Y.}~\bibnamefont {Murashita}}, \bibinfo {author} {\bibfnamefont {A.}~\bibnamefont {Noguchi}}, \bibinfo {author} {\bibfnamefont {S.}~\bibnamefont {Kono}}, \bibinfo {author} {\bibfnamefont {Y.}~\bibnamefont {Tabuchi}}, \bibinfo {author} {\bibfnamefont {R.}~\bibnamefont {Yamazaki}}, \bibinfo {author} {\bibfnamefont {M.}~\bibnamefont {Ueda}},\ and\ \bibinfo {author} {\bibfnamefont {Y.}~\bibnamefont {Nakamura}},\ }\bibfield  {title} {\bibinfo {title} {Information-to-work conversion by {M}axwell's demon in a superconducting circuit quantum electrodynamical system},\ }\href@noop {} {\bibfield  {journal} {\bibinfo  {journal} {Nature Comm.}\ }\textbf {\bibinfo {volume} {9}},\ \bibinfo {pages} {1291} (\bibinfo {year} {2018})}\BibitemShut {NoStop}%
\bibitem [{\citenamefont {Naghiloo}\ \emph {et~al.}(2018)\citenamefont {Naghiloo}, \citenamefont {Alonso}, \citenamefont {Romito}, \citenamefont {Lutz},\ and\ \citenamefont {Murch}}]{nag18}%
  \BibitemOpen
  \bibfield  {author} {\bibinfo {author} {\bibfnamefont {M.}~\bibnamefont {Naghiloo}}, \bibinfo {author} {\bibfnamefont {J.~J.}\ \bibnamefont {Alonso}}, \bibinfo {author} {\bibfnamefont {A.}~\bibnamefont {Romito}}, \bibinfo {author} {\bibfnamefont {E.}~\bibnamefont {Lutz}},\ and\ \bibinfo {author} {\bibfnamefont {K.~W.}\ \bibnamefont {Murch}},\ }\bibfield  {title} {\bibinfo {title} {Information gain and loss for a quantum {M}axwell's demon},\ }\href@noop {} {\bibfield  {journal} {\bibinfo  {journal} {Phys. Rev. Lett.}\ }\textbf {\bibinfo {volume} {121}},\ \bibinfo {pages} {030604} (\bibinfo {year} {2018})}\BibitemShut {NoStop}%
\bibitem [{\citenamefont {Najera-Santos}\ \emph {et~al.}(2020)\citenamefont {Najera-Santos}, \citenamefont {Camati}, \citenamefont {M\'etillon}, \citenamefont {Brune}, \citenamefont {Raimond}, \citenamefont {Auff\`eves},\ and\ \citenamefont {Dotsenko}}]{naj20}%
  \BibitemOpen
  \bibfield  {author} {\bibinfo {author} {\bibfnamefont {B.~L.}\ \bibnamefont {Najera-Santos}}, \bibinfo {author} {\bibfnamefont {P.~A.}\ \bibnamefont {Camati}}, \bibinfo {author} {\bibfnamefont {V.}~\bibnamefont {M\'etillon}}, \bibinfo {author} {\bibfnamefont {M.}~\bibnamefont {Brune}}, \bibinfo {author} {\bibfnamefont {J.~M.}\ \bibnamefont {Raimond}}, \bibinfo {author} {\bibfnamefont {A.}~\bibnamefont {Auff\`eves}},\ and\ \bibinfo {author} {\bibfnamefont {I.}~\bibnamefont {Dotsenko}},\ }\bibfield  {title} {\bibinfo {title} {Autonomous {M}axwell's demon in a cavity {QED} system},\ }\href@noop {} {\bibfield  {journal} {\bibinfo  {journal} {Phys. Rev. Research}\ }\textbf {\bibinfo {volume} {2}},\ \bibinfo {pages} {032025} (\bibinfo {year} {2020})}\BibitemShut {NoStop}%
\bibitem [{\citenamefont {Auletta}\ \emph {et~al.}(2009)\citenamefont {Auletta}, \citenamefont {Fortunato},\ and\ \citenamefont {Parisi}}]{aul09}%
  \BibitemOpen
  \bibfield  {author} {\bibinfo {author} {\bibfnamefont {G.}~\bibnamefont {Auletta}}, \bibinfo {author} {\bibfnamefont {M.}~\bibnamefont {Fortunato}},\ and\ \bibinfo {author} {\bibfnamefont {G.}~\bibnamefont {Parisi}},\ }\href@noop {} {\emph {\bibinfo {title} {Quantum Mechanics}}}\ (\bibinfo  {publisher} {Cambridge University Press, Cambridge},\ \bibinfo {year} {2009})\BibitemShut {NoStop}%
\bibitem [{\citenamefont {Kim}\ \emph {et~al.}(2011)\citenamefont {Kim}, \citenamefont {Sagawa}, \citenamefont {De~Liberato},\ and\ \citenamefont {Ueda}}]{kim11}%
  \BibitemOpen
  \bibfield  {author} {\bibinfo {author} {\bibfnamefont {S.~W.}\ \bibnamefont {Kim}}, \bibinfo {author} {\bibfnamefont {T.}~\bibnamefont {Sagawa}}, \bibinfo {author} {\bibfnamefont {S.}~\bibnamefont {De~Liberato}},\ and\ \bibinfo {author} {\bibfnamefont {M.}~\bibnamefont {Ueda}},\ }\bibfield  {title} {\bibinfo {title} {Quantum {S}zilard engine},\ }\href@noop {} {\bibfield  {journal} {\bibinfo  {journal} {Phys. Rev. Lett.}\ }\textbf {\bibinfo {volume} {106}},\ \bibinfo {pages} {070401} (\bibinfo {year} {2011})}\BibitemShut {NoStop}%
\bibitem [{\citenamefont {Cai}\ \emph {et~al.}(2012)\citenamefont {Cai}, \citenamefont {Dong},\ and\ \citenamefont {Sun}}]{cai12}%
  \BibitemOpen
  \bibfield  {author} {\bibinfo {author} {\bibfnamefont {C.~Y.}\ \bibnamefont {Cai}}, \bibinfo {author} {\bibfnamefont {H.}~\bibnamefont {Dong}},\ and\ \bibinfo {author} {\bibfnamefont {C.~P.}\ \bibnamefont {Sun}},\ }\bibfield  {title} {\bibinfo {title} {Multiparticle quantum {S}zilard engine with optimal cycles assisted by a maxwell's demon},\ }\href@noop {} {\bibfield  {journal} {\bibinfo  {journal} {Phys. Rev. E}\ }\textbf {\bibinfo {volume} {85}},\ \bibinfo {pages} {031114} (\bibinfo {year} {2012})}\BibitemShut {NoStop}%
\bibitem [{\citenamefont {Jeon}\ and\ \citenamefont {Kim}(2016)}]{jeo16}%
  \BibitemOpen
  \bibfield  {author} {\bibinfo {author} {\bibfnamefont {H.~J.}\ \bibnamefont {Jeon}}\ and\ \bibinfo {author} {\bibfnamefont {S.~W.}\ \bibnamefont {Kim}},\ }\bibfield  {title} {\bibinfo {title} {Optimal work of the quantum {S}zilard engine under isothermal processes with inevitable irreversibility},\ }\href@noop {} {\bibfield  {journal} {\bibinfo  {journal} {New J. Phys.}\ }\textbf {\bibinfo {volume} {18}},\ \bibinfo {pages} {043002} (\bibinfo {year} {2016})}\BibitemShut {NoStop}%
\bibitem [{\citenamefont {Bengtsson}\ \emph {et~al.}(2018{\natexlab{a}})\citenamefont {Bengtsson}, \citenamefont {Tengstrand}, \citenamefont {Wacker}, \citenamefont {Samuelsson}, \citenamefont {Ueda}, \citenamefont {Linke},\ and\ \citenamefont {Reimann}}]{ben18}%
  \BibitemOpen
  \bibfield  {author} {\bibinfo {author} {\bibfnamefont {J.}~\bibnamefont {Bengtsson}}, \bibinfo {author} {\bibfnamefont {M.~N.}\ \bibnamefont {Tengstrand}}, \bibinfo {author} {\bibfnamefont {A.}~\bibnamefont {Wacker}}, \bibinfo {author} {\bibfnamefont {P.}~\bibnamefont {Samuelsson}}, \bibinfo {author} {\bibfnamefont {M.}~\bibnamefont {Ueda}}, \bibinfo {author} {\bibfnamefont {H.}~\bibnamefont {Linke}},\ and\ \bibinfo {author} {\bibfnamefont {S.~M.}\ \bibnamefont {Reimann}},\ }\bibfield  {title} {\bibinfo {title} {Quantum {S}zilard engine with attractively interacting bosons},\ }\href@noop {} {\bibfield  {journal} {\bibinfo  {journal} {Phys. Rev. Lett.}\ }\textbf {\bibinfo {volume} {120}},\ \bibinfo {pages} {100601} (\bibinfo {year} {2018}{\natexlab{a}})}\BibitemShut {NoStop}%
\bibitem [{\citenamefont {Bengtsson}\ \emph {et~al.}(2018{\natexlab{b}})\citenamefont {Bengtsson}, \citenamefont {Tengstrand},\ and\ \citenamefont {Reimann}}]{ben18a}%
  \BibitemOpen
  \bibfield  {author} {\bibinfo {author} {\bibfnamefont {J.}~\bibnamefont {Bengtsson}}, \bibinfo {author} {\bibfnamefont {M.~N.}\ \bibnamefont {Tengstrand}},\ and\ \bibinfo {author} {\bibfnamefont {S.~M.}\ \bibnamefont {Reimann}},\ }\bibfield  {title} {\bibinfo {title} {Bosonic {S}zilard engine assisted by {F}eshbach resonances},\ }\href@noop {} {\bibfield  {journal} {\bibinfo  {journal} {Phys. Rev. A}\ }\textbf {\bibinfo {volume} {97}},\ \bibinfo {pages} {062128} (\bibinfo {year} {2018}{\natexlab{b}})}\BibitemShut {NoStop}%
\bibitem [{\citenamefont {Plesch}\ \emph {et~al.}(2013)\citenamefont {Plesch}, \citenamefont {Dahlsten}, \citenamefont {Goold},\ and\ \citenamefont {Vedral}}]{ple13}%
  \BibitemOpen
  \bibfield  {author} {\bibinfo {author} {\bibfnamefont {M.}~\bibnamefont {Plesch}}, \bibinfo {author} {\bibfnamefont {O.}~\bibnamefont {Dahlsten}}, \bibinfo {author} {\bibfnamefont {J.}~\bibnamefont {Goold}},\ and\ \bibinfo {author} {\bibfnamefont {V.}~\bibnamefont {Vedral}},\ }\bibfield  {title} {\bibinfo {title} {Comment on "quantum {S}zilard engine''},\ }\href@noop {} {\bibfield  {journal} {\bibinfo  {journal} {Phys. Rev. Lett.}\ }\textbf {\bibinfo {volume} {111}},\ \bibinfo {pages} {188901} (\bibinfo {year} {2013})}\BibitemShut {NoStop}%
\bibitem [{\citenamefont {Kim}\ \emph {et~al.}(2013)\citenamefont {Kim}, \citenamefont {Kim}, \citenamefont {Sagawa}, \citenamefont {De~Liberato},\ and\ \citenamefont {Ueda}}]{kim13}%
  \BibitemOpen
  \bibfield  {author} {\bibinfo {author} {\bibfnamefont {S.~W.}\ \bibnamefont {Kim}}, \bibinfo {author} {\bibfnamefont {K.-H.}\ \bibnamefont {Kim}}, \bibinfo {author} {\bibfnamefont {T.}~\bibnamefont {Sagawa}}, \bibinfo {author} {\bibfnamefont {S.}~\bibnamefont {De~Liberato}},\ and\ \bibinfo {author} {\bibfnamefont {M.}~\bibnamefont {Ueda}},\ }\bibfield  {title} {\bibinfo {title} {Kim et al. reply:},\ }\href@noop {} {\bibfield  {journal} {\bibinfo  {journal} {Phys. Rev. Lett.}\ }\textbf {\bibinfo {volume} {111}},\ \bibinfo {pages} {188902} (\bibinfo {year} {2013})}\BibitemShut {NoStop}%
\bibitem [{\citenamefont {Plesch}\ \emph {et~al.}(2014)\citenamefont {Plesch}, \citenamefont {Dahlsten}, \citenamefont {Goold},\ and\ \citenamefont {Vedral}}]{ple14}%
  \BibitemOpen
  \bibfield  {author} {\bibinfo {author} {\bibfnamefont {M.}~\bibnamefont {Plesch}}, \bibinfo {author} {\bibfnamefont {O.}~\bibnamefont {Dahlsten}}, \bibinfo {author} {\bibfnamefont {J.}~\bibnamefont {Goold}},\ and\ \bibinfo {author} {\bibfnamefont {V.}~\bibnamefont {Vedral}},\ }\bibfield  {title} {\bibinfo {title} {Maxwell's daemon: Information versus particle statistics},\ }\href@noop {} {\bibfield  {journal} {\bibinfo  {journal} {Sci. Rep.}\ }\textbf {\bibinfo {volume} {4}},\ \bibinfo {pages} {6995} (\bibinfo {year} {2014})}\BibitemShut {NoStop}%
\bibitem [{\citenamefont {Holmes}\ \emph {et~al.}(2020)\citenamefont {Holmes}, \citenamefont {Anders},\ and\ \citenamefont {Mintert}}]{hol20}%
  \BibitemOpen
  \bibfield  {author} {\bibinfo {author} {\bibfnamefont {Z.}~\bibnamefont {Holmes}}, \bibinfo {author} {\bibfnamefont {J.}~\bibnamefont {Anders}},\ and\ \bibinfo {author} {\bibfnamefont {F.}~\bibnamefont {Mintert}},\ }\bibfield  {title} {\bibinfo {title} {Enhanced energy transfer to an optomechanical piston from indistinguishable photons},\ }\href@noop {} {\bibfield  {journal} {\bibinfo  {journal} {Phys. Rev. Lett.}\ }\textbf {\bibinfo {volume} {124}},\ \bibinfo {pages} {210601} (\bibinfo {year} {2020})}\BibitemShut {NoStop}%
\bibitem [{\citenamefont {Myers}\ and\ \citenamefont {Deffner}(2020)}]{mye20}%
  \BibitemOpen
  \bibfield  {author} {\bibinfo {author} {\bibfnamefont {N.~M.}\ \bibnamefont {Myers}}\ and\ \bibinfo {author} {\bibfnamefont {S.}~\bibnamefont {Deffner}},\ }\bibfield  {title} {\bibinfo {title} {Bosons outperform fermions: The thermodynamic advantage of symmetry},\ }\href@noop {} {\bibfield  {journal} {\bibinfo  {journal} {Phys. Rev. E}\ }\textbf {\bibinfo {volume} {101}},\ \bibinfo {pages} {012110} (\bibinfo {year} {2020})}\BibitemShut {NoStop}%
\bibitem [{\citenamefont {Watanabe}\ \emph {et~al.}(2020)\citenamefont {Watanabe}, \citenamefont {Venkatesh}, \citenamefont {Talkner}, \citenamefont {Hwang},\ and\ \citenamefont {del Campo}}]{wat20}%
  \BibitemOpen
  \bibfield  {author} {\bibinfo {author} {\bibfnamefont {G.}~\bibnamefont {Watanabe}}, \bibinfo {author} {\bibfnamefont {B.~P.}\ \bibnamefont {Venkatesh}}, \bibinfo {author} {\bibfnamefont {P.}~\bibnamefont {Talkner}}, \bibinfo {author} {\bibfnamefont {M.-J.}\ \bibnamefont {Hwang}},\ and\ \bibinfo {author} {\bibfnamefont {A.}~\bibnamefont {del Campo}},\ }\bibfield  {title} {\bibinfo {title} {Quantum statistical enhancement of the collective performance of multiple bosonic engines},\ }\href@noop {} {\bibfield  {journal} {\bibinfo  {journal} {Phys. Rev. Lett.}\ }\textbf {\bibinfo {volume} {124}},\ \bibinfo {pages} {210603} (\bibinfo {year} {2020})}\BibitemShut {NoStop}%
\bibitem [{\citenamefont {Zanin}\ \emph {et~al.}(2022)\citenamefont {Zanin}, \citenamefont {Antesberger}, \citenamefont {Jacquet}, \citenamefont {Ribeiro}, \citenamefont {Rozema},\ and\ \citenamefont {Walther}}]{zan22}%
  \BibitemOpen
  \bibfield  {author} {\bibinfo {author} {\bibfnamefont {G.~L.}\ \bibnamefont {Zanin}}, \bibinfo {author} {\bibfnamefont {M.}~\bibnamefont {Antesberger}}, \bibinfo {author} {\bibfnamefont {M.~J.}\ \bibnamefont {Jacquet}}, \bibinfo {author} {\bibfnamefont {P.~H.~S.}\ \bibnamefont {Ribeiro}}, \bibinfo {author} {\bibfnamefont {L.~A.}\ \bibnamefont {Rozema}},\ and\ \bibinfo {author} {\bibfnamefont {P.}~\bibnamefont {Walther}},\ }\bibfield  {title} {\bibinfo {title} {Enhanced photonic {M}axwellÕs demon with correlated baths},\ }\href@noop {} {\bibfield  {journal} {\bibinfo  {journal} {Quantum}\ }\textbf {\bibinfo {volume} {6}},\ \bibinfo {pages} {810} (\bibinfo {year} {2022})}\BibitemShut {NoStop}%
\bibitem [{\citenamefont {Vidrighin}\ \emph {et~al.}(2016)\citenamefont {Vidrighin}, \citenamefont {Dahlsten}, \citenamefont {Barbieri}, \citenamefont {Kim}, \citenamefont {Vedral},\ and\ \citenamefont {Walmsley}}]{vid16}%
  \BibitemOpen
  \bibfield  {author} {\bibinfo {author} {\bibfnamefont {M.~D.}\ \bibnamefont {Vidrighin}}, \bibinfo {author} {\bibfnamefont {O.}~\bibnamefont {Dahlsten}}, \bibinfo {author} {\bibfnamefont {M.}~\bibnamefont {Barbieri}}, \bibinfo {author} {\bibfnamefont {M.~S.}\ \bibnamefont {Kim}}, \bibinfo {author} {\bibfnamefont {V.}~\bibnamefont {Vedral}},\ and\ \bibinfo {author} {\bibfnamefont {I.~A.}\ \bibnamefont {Walmsley}},\ }\bibfield  {title} {\bibinfo {title} {Photonic {M}axwell's demon},\ }\href@noop {} {\bibfield  {journal} {\bibinfo  {journal} {Phys. Rev. Lett.}\ }\textbf {\bibinfo {volume} {116}},\ \bibinfo {pages} {050401} (\bibinfo {year} {2016})}\BibitemShut {NoStop}%
\bibitem [{\citenamefont {Neergaard-Nielsen}\ \emph {et~al.}(2006)\citenamefont {Neergaard-Nielsen}, \citenamefont {Nielsen}, \citenamefont {Hettich}, \citenamefont {M\o{}lmer},\ and\ \citenamefont {Polzik}}]{nee06}%
  \BibitemOpen
  \bibfield  {author} {\bibinfo {author} {\bibfnamefont {J.~S.}\ \bibnamefont {Neergaard-Nielsen}}, \bibinfo {author} {\bibfnamefont {B.~M.}\ \bibnamefont {Nielsen}}, \bibinfo {author} {\bibfnamefont {C.}~\bibnamefont {Hettich}}, \bibinfo {author} {\bibfnamefont {K.}~\bibnamefont {M\o{}lmer}},\ and\ \bibinfo {author} {\bibfnamefont {E.~S.}\ \bibnamefont {Polzik}},\ }\bibfield  {title} {\bibinfo {title} {Generation of a superposition of odd photon number states for quantum information networks},\ }\href@noop {} {\bibfield  {journal} {\bibinfo  {journal} {Phys. Rev. Lett.}\ }\textbf {\bibinfo {volume} {97}},\ \bibinfo {pages} {083604} (\bibinfo {year} {2006})}\BibitemShut {NoStop}%
\bibitem [{\citenamefont {Ourjoumtsev}\ \emph {et~al.}(2006)\citenamefont {Ourjoumtsev}, \citenamefont {Tualle-Brouri}, \citenamefont {Laurat},\ and\ \citenamefont {Grangier}}]{our06}%
  \BibitemOpen
  \bibfield  {author} {\bibinfo {author} {\bibfnamefont {A.}~\bibnamefont {Ourjoumtsev}}, \bibinfo {author} {\bibfnamefont {R.}~\bibnamefont {Tualle-Brouri}}, \bibinfo {author} {\bibfnamefont {J.}~\bibnamefont {Laurat}},\ and\ \bibinfo {author} {\bibfnamefont {P.}~\bibnamefont {Grangier}},\ }\bibfield  {title} {\bibinfo {title} {Generating optical {S}chr\"odinger kittens for quantum information processing},\ }\href@noop {} {\bibfield  {journal} {\bibinfo  {journal} {Science}\ }\textbf {\bibinfo {volume} {312}},\ \bibinfo {pages} {83} (\bibinfo {year} {2006})}\BibitemShut {NoStop}%
\bibitem [{\citenamefont {Parigi}\ \emph {et~al.}(2007)\citenamefont {Parigi}, \citenamefont {Zavatta}, \citenamefont {Kim},\ and\ \citenamefont {Bellini}}]{par07}%
  \BibitemOpen
  \bibfield  {author} {\bibinfo {author} {\bibfnamefont {V.}~\bibnamefont {Parigi}}, \bibinfo {author} {\bibfnamefont {A.}~\bibnamefont {Zavatta}}, \bibinfo {author} {\bibfnamefont {M.}~\bibnamefont {Kim}},\ and\ \bibinfo {author} {\bibfnamefont {M.}~\bibnamefont {Bellini}},\ }\bibfield  {title} {\bibinfo {title} {Probing quantum commutation rules by addition and subtraction of single photons to/from a light field},\ }\href@noop {} {\bibfield  {journal} {\bibinfo  {journal} {Science}\ }\textbf {\bibinfo {volume} {317}},\ \bibinfo {pages} {1890} (\bibinfo {year} {2007})}\BibitemShut {NoStop}%
\bibitem [{\citenamefont {Zavatta}\ \emph {et~al.}(2008)\citenamefont {Zavatta}, \citenamefont {Parigi}, \citenamefont {Kim},\ and\ \citenamefont {Bellini}}]{Zavatta2008}%
  \BibitemOpen
  \bibfield  {author} {\bibinfo {author} {\bibfnamefont {A.}~\bibnamefont {Zavatta}}, \bibinfo {author} {\bibfnamefont {V.}~\bibnamefont {Parigi}}, \bibinfo {author} {\bibfnamefont {M.~S.}\ \bibnamefont {Kim}},\ and\ \bibinfo {author} {\bibfnamefont {M.}~\bibnamefont {Bellini}},\ }\bibfield  {title} {\bibinfo {title} {Subtracting photons from arbitrary light fields: experimental test of coherent state invariance by single-photon annihilation},\ }\href@noop {} {\bibfield  {journal} {\bibinfo  {journal} {New J. Phys.}\ }\textbf {\bibinfo {volume} {10}},\ \bibinfo {pages} {123006} (\bibinfo {year} {2008})}\BibitemShut {NoStop}%
\bibitem [{\citenamefont {Fedorov}\ \emph {et~al.}(2015)\citenamefont {Fedorov}, \citenamefont {Ulanov}, \citenamefont {Kurochkin},\ and\ \citenamefont {Lvovsky}}]{Fedorov2015}%
  \BibitemOpen
  \bibfield  {author} {\bibinfo {author} {\bibfnamefont {I.~A.}\ \bibnamefont {Fedorov}}, \bibinfo {author} {\bibfnamefont {A.~E.}\ \bibnamefont {Ulanov}}, \bibinfo {author} {\bibfnamefont {Y.~V.}\ \bibnamefont {Kurochkin}},\ and\ \bibinfo {author} {\bibfnamefont {A.~I.}\ \bibnamefont {Lvovsky}},\ }\bibfield  {title} {\bibinfo {title} {Quantum vampire: collapse-free action at a distance by the photon annihilation operator},\ }\href@noop {} {\bibfield  {journal} {\bibinfo  {journal} {Optica}\ }\textbf {\bibinfo {volume} {2}},\ \bibinfo {pages} {112} (\bibinfo {year} {2015})}\BibitemShut {NoStop}%
\bibitem [{\citenamefont {Bogdanov}\ \emph {et~al.}(2017)\citenamefont {Bogdanov}, \citenamefont {Katamadze}, \citenamefont {Avosopiants}, \citenamefont {Belinsky}, \citenamefont {Bogdanova}, \citenamefont {Kalinkin},\ and\ \citenamefont {Kulik}}]{Bogdanov2017}%
  \BibitemOpen
  \bibfield  {author} {\bibinfo {author} {\bibfnamefont {Y.~I.}\ \bibnamefont {Bogdanov}}, \bibinfo {author} {\bibfnamefont {K.~G.}\ \bibnamefont {Katamadze}}, \bibinfo {author} {\bibfnamefont {G.~V.}\ \bibnamefont {Avosopiants}}, \bibinfo {author} {\bibfnamefont {L.~V.}\ \bibnamefont {Belinsky}}, \bibinfo {author} {\bibfnamefont {N.~A.}\ \bibnamefont {Bogdanova}}, \bibinfo {author} {\bibfnamefont {A.~A.}\ \bibnamefont {Kalinkin}},\ and\ \bibinfo {author} {\bibfnamefont {S.~P.}\ \bibnamefont {Kulik}},\ }\bibfield  {title} {\bibinfo {title} {Multiphoton subtracted thermal states: Description, preparation, and reconstruction},\ }\href@noop {} {\bibfield  {journal} {\bibinfo  {journal} {Phys. Rev. A}\ }\textbf {\bibinfo {volume} {96}} (\bibinfo {year} {2017})}\BibitemShut {NoStop}%
\bibitem [{\citenamefont {Hlou{\v{s}}ek}\ \emph {et~al.}(2017)\citenamefont {Hlou{\v{s}}ek}, \citenamefont {Je{\v{z}}ek},\ and\ \citenamefont {Filip}}]{Hlousek2017}%
  \BibitemOpen
  \bibfield  {author} {\bibinfo {author} {\bibfnamefont {J.}~\bibnamefont {Hlou{\v{s}}ek}}, \bibinfo {author} {\bibfnamefont {M.}~\bibnamefont {Je{\v{z}}ek}},\ and\ \bibinfo {author} {\bibfnamefont {R.}~\bibnamefont {Filip}},\ }\bibfield  {title} {\bibinfo {title} {Work and information from thermal states after subtraction of energy quanta},\ }\href@noop {} {\bibfield  {journal} {\bibinfo  {journal} {Sci. Rep.}\ }\textbf {\bibinfo {volume} {7}},\ \bibinfo {pages} {13046} (\bibinfo {year} {2017})}\BibitemShut {NoStop}%
\bibitem [{\citenamefont {Rafsanjani}\ \emph {et~al.}(2017)\citenamefont {Rafsanjani}, \citenamefont {Mirhosseini}, \citenamefont {Maga{\~{n}}a-Loaiza}, \citenamefont {Gard}, \citenamefont {Birrittella}, \citenamefont {Koltenbah}, \citenamefont {Parazzoli}, \citenamefont {Capron}, \citenamefont {Gerry}, \citenamefont {Dowling},\ and\ \citenamefont {Boyd}}]{HashemiRafsanjani2017}%
  \BibitemOpen
  \bibfield  {author} {\bibinfo {author} {\bibfnamefont {S.~M.~H.}\ \bibnamefont {Rafsanjani}}, \bibinfo {author} {\bibfnamefont {M.}~\bibnamefont {Mirhosseini}}, \bibinfo {author} {\bibfnamefont {O.~S.}\ \bibnamefont {Maga{\~{n}}a-Loaiza}}, \bibinfo {author} {\bibfnamefont {B.~T.}\ \bibnamefont {Gard}}, \bibinfo {author} {\bibfnamefont {R.}~\bibnamefont {Birrittella}}, \bibinfo {author} {\bibfnamefont {B.~E.}\ \bibnamefont {Koltenbah}}, \bibinfo {author} {\bibfnamefont {C.~G.}\ \bibnamefont {Parazzoli}}, \bibinfo {author} {\bibfnamefont {B.~A.}\ \bibnamefont {Capron}}, \bibinfo {author} {\bibfnamefont {C.~C.}\ \bibnamefont {Gerry}}, \bibinfo {author} {\bibfnamefont {J.~P.}\ \bibnamefont {Dowling}},\ and\ \bibinfo {author} {\bibfnamefont {R.}~\bibnamefont {Boyd}},\ }\bibfield  {title} {\bibinfo {title} {Quantum-enhanced interferometry with weak thermal light},\ }\href@noop {} {\bibfield  {journal} {\bibinfo  {journal} {Optica}\ }\textbf {\bibinfo {volume} {4}},\ \bibinfo {pages} {487} (\bibinfo {year}
  {2017})}\BibitemShut {NoStop}%
\bibitem [{\citenamefont {Katamadze}\ \emph {et~al.}(2018)\citenamefont {Katamadze}, \citenamefont {Avosopiants}, \citenamefont {Bogdanov},\ and\ \citenamefont {Kulik}}]{Katamadze2018}%
  \BibitemOpen
  \bibfield  {author} {\bibinfo {author} {\bibfnamefont {K.~G.}\ \bibnamefont {Katamadze}}, \bibinfo {author} {\bibfnamefont {G.~V.}\ \bibnamefont {Avosopiants}}, \bibinfo {author} {\bibfnamefont {Y.~I.}\ \bibnamefont {Bogdanov}},\ and\ \bibinfo {author} {\bibfnamefont {S.~P.}\ \bibnamefont {Kulik}},\ }\bibfield  {title} {\bibinfo {title} {How quantum is the {\textquotedblleft}quantum vampire{\textquotedblright} effect?: testing with thermal light},\ }\href@noop {} {\bibfield  {journal} {\bibinfo  {journal} {Optica}\ }\textbf {\bibinfo {volume} {5}},\ \bibinfo {pages} {723} (\bibinfo {year} {2018})}\BibitemShut {NoStop}%
\bibitem [{\citenamefont {Loudon}(2010)}]{Loudon2010}%
  \BibitemOpen
  \bibfield  {author} {\bibinfo {author} {\bibfnamefont {R.}~\bibnamefont {Loudon}},\ }\href@noop {} {\emph {\bibinfo {title} {The Quantum Theory of Light}}}\ (\bibinfo  {publisher} {Oxford University Press, Oxford},\ \bibinfo {year} {2010})\BibitemShut {NoStop}%
\bibitem [{\citenamefont {Pietzonka}\ and\ \citenamefont {Seifert}(2018)}]{pie18}%
  \BibitemOpen
  \bibfield  {author} {\bibinfo {author} {\bibfnamefont {P.}~\bibnamefont {Pietzonka}}\ and\ \bibinfo {author} {\bibfnamefont {U.}~\bibnamefont {Seifert}},\ }\bibfield  {title} {\bibinfo {title} {Universal trade-off between power, efficiency, and constancy in steady-state heat engines},\ }\href@noop {} {\bibfield  {journal} {\bibinfo  {journal} {Phys. Rev. Lett.}\ }\textbf {\bibinfo {volume} {120}},\ \bibinfo {pages} {190602} (\bibinfo {year} {2018})}\BibitemShut {NoStop}%
\bibitem [{\citenamefont {Holubec}\ and\ \citenamefont {Ryabov}(2018)}]{hol18}%
  \BibitemOpen
  \bibfield  {author} {\bibinfo {author} {\bibfnamefont {V.}~\bibnamefont {Holubec}}\ and\ \bibinfo {author} {\bibfnamefont {A.}~\bibnamefont {Ryabov}},\ }\bibfield  {title} {\bibinfo {title} {Cycling tames power fluctuations near optimum efficiency},\ }\href@noop {} {\bibfield  {journal} {\bibinfo  {journal} {Phys. Rev. Lett.}\ }\textbf {\bibinfo {volume} {121}},\ \bibinfo {pages} {120601} (\bibinfo {year} {2018})}\BibitemShut {NoStop}%
\bibitem [{\citenamefont {Denzler}\ and\ \citenamefont {Lutz}(2021)}]{den21}%
  \BibitemOpen
  \bibfield  {author} {\bibinfo {author} {\bibfnamefont {T.}~\bibnamefont {Denzler}}\ and\ \bibinfo {author} {\bibfnamefont {E.}~\bibnamefont {Lutz}},\ }\bibfield  {title} {\bibinfo {title} {Power fluctuations in a finite-time quantum carnot engine},\ }\href@noop {} {\bibfield  {journal} {\bibinfo  {journal} {Phys. Rev. Research}\ }\textbf {\bibinfo {volume} {3}},\ \bibinfo {pages} {L032041} (\bibinfo {year} {2021})}\BibitemShut {NoStop}%
\bibitem [{\citenamefont {Partovi}(1989)}]{par89}%
  \BibitemOpen
  \bibfield  {author} {\bibinfo {author} {\bibfnamefont {M.~H.}\ \bibnamefont {Partovi}},\ }\bibfield  {title} {\bibinfo {title} {Quantum thermodynamics},\ }\href@noop {} {\bibfield  {journal} {\bibinfo  {journal} {Phys. Lett. A}\ }\textbf {\bibinfo {volume} {137}},\ \bibinfo {pages} {440} (\bibinfo {year} {1989})}\BibitemShut {NoStop}%
\bibitem [{\citenamefont {Peres}(2002)}]{per02}%
  \BibitemOpen
  \bibfield  {author} {\bibinfo {author} {\bibfnamefont {A.}~\bibnamefont {Peres}},\ }\href@noop {} {\emph {\bibinfo {title} {Quantum Theory: {C}oncepts and Methods}}}\ (\bibinfo  {publisher} {Kluwer Academic Publishers, New York},\ \bibinfo {year} {2002})\BibitemShut {NoStop}%
\bibitem [{\citenamefont {Gaveau}\ \emph {et~al.}(2014)\citenamefont {Gaveau}, \citenamefont {Granger}, \citenamefont {Moreau},\ and\ \citenamefont {Schulman}}]{gav14}%
  \BibitemOpen
  \bibfield  {author} {\bibinfo {author} {\bibfnamefont {B.}~\bibnamefont {Gaveau}}, \bibinfo {author} {\bibfnamefont {L.}~\bibnamefont {Granger}}, \bibinfo {author} {\bibfnamefont {M.}~\bibnamefont {Moreau}},\ and\ \bibinfo {author} {\bibfnamefont {L.~S.}\ \bibnamefont {Schulman}},\ }\bibfield  {title} {\bibinfo {title} {Relative entropy, interaction energy and the nature of dissipation},\ }\href {https://doi.org/10.3390/e16063173} {\bibfield  {journal} {\bibinfo  {journal} {Entropy}\ }\textbf {\bibinfo {volume} {16}},\ \bibinfo {pages} {3173} (\bibinfo {year} {2014})}\BibitemShut {NoStop}%
\bibitem [{\citenamefont {Mayer}\ \emph {et~al.}(2023)\citenamefont {Mayer}, \citenamefont {Lutz},\ and\ \citenamefont {Widera}}]{may23}%
  \BibitemOpen
  \bibfield  {author} {\bibinfo {author} {\bibfnamefont {D.}~\bibnamefont {Mayer}}, \bibinfo {author} {\bibfnamefont {E.}~\bibnamefont {Lutz}},\ and\ \bibinfo {author} {\bibfnamefont {A.}~\bibnamefont {Widera}},\ }\bibfield  {title} {\bibinfo {title} {Generalized {C}lausius inequalities in a nonequilibrium cold-atom system},\ }\href@noop {} {\bibfield  {journal} {\bibinfo  {journal} {Commun. Phys.}\ }\textbf {\bibinfo {volume} {6}},\ \bibinfo {pages} {61} (\bibinfo {year} {2023})}\BibitemShut {NoStop}%
\bibitem [{sup()}]{sup}%
  \BibitemOpen
  \href@noop {} {}\bibinfo {note} {See Supplemental Material}\BibitemShut {NoStop}%
\bibitem [{\citenamefont {\v{S}varc}\ \emph {et~al.}(2019)\citenamefont {\v{S}varc}, \citenamefont {Nov\'{a}kov\'{a}}, \citenamefont {Mazin},\ and\ \citenamefont {Je\v{z}ek}}]{Svarc2019}%
  \BibitemOpen
  \bibfield  {author} {\bibinfo {author} {\bibfnamefont {V.}~\bibnamefont {\v{S}varc}}, \bibinfo {author} {\bibfnamefont {M.}~\bibnamefont {Nov\'{a}kov\'{a}}}, \bibinfo {author} {\bibfnamefont {G.}~\bibnamefont {Mazin}},\ and\ \bibinfo {author} {\bibfnamefont {M.}~\bibnamefont {Je\v{z}ek}},\ }\bibfield  {title} {\bibinfo {title} {Fully tunable and switchable coupler for photonic routing in quantum detection and modulation},\ }\href {https://doi.org/10.1364/OL.44.005844} {\bibfield  {journal} {\bibinfo  {journal} {Opt. Lett.}\ }\textbf {\bibinfo {volume} {44}},\ \bibinfo {pages} {5844} (\bibinfo {year} {2019})}\BibitemShut {NoStop}%
\bibitem [{\citenamefont {Hlou{\v{s}}ek}\ \emph {et~al.}(2019)\citenamefont {Hlou{\v{s}}ek}, \citenamefont {Dudka}, \citenamefont {Straka},\ and\ \citenamefont {Je{\v{z}}ek}}]{Hlousek2019}%
  \BibitemOpen
  \bibfield  {author} {\bibinfo {author} {\bibfnamefont {J.}~\bibnamefont {Hlou{\v{s}}ek}}, \bibinfo {author} {\bibfnamefont {M.}~\bibnamefont {Dudka}}, \bibinfo {author} {\bibfnamefont {I.}~\bibnamefont {Straka}},\ and\ \bibinfo {author} {\bibfnamefont {M.}~\bibnamefont {Je{\v{z}}ek}},\ }\bibfield  {title} {\bibinfo {title} {Accurate detection of arbitrary photon statistics},\ }\href@noop {} {\bibfield  {journal} {\bibinfo  {journal} {Phys. Rev. Lett.}\ }\textbf {\bibinfo {volume} {123}},\ \bibinfo {pages} {153604} (\bibinfo {year} {2019})}\BibitemShut {NoStop}%
\bibitem [{\citenamefont {Cover}\ and\ \citenamefont {Thomas}(2006)}]{cov06}%
  \BibitemOpen
  \bibfield  {author} {\bibinfo {author} {\bibfnamefont {T.~M.}\ \bibnamefont {Cover}}\ and\ \bibinfo {author} {\bibfnamefont {J.~A.}\ \bibnamefont {Thomas}},\ }\href@noop {} {\emph {\bibinfo {title} {Elements of Information Theory}}}\ (\bibinfo  {publisher} {Wiley, New York},\ \bibinfo {year} {2006})\BibitemShut {NoStop}%
\bibitem [{\citenamefont {Deffner}\ and\ \citenamefont {Lutz}(2011)}]{def11}%
  \BibitemOpen
  \bibfield  {author} {\bibinfo {author} {\bibfnamefont {S.}~\bibnamefont {Deffner}}\ and\ \bibinfo {author} {\bibfnamefont {E.}~\bibnamefont {Lutz}},\ }\bibfield  {title} {\bibinfo {title} {Nonequilibrium entropy production for open quantum systems},\ }\href@noop {} {\bibfield  {journal} {\bibinfo  {journal} {Phys. Rev. Lett.}\ }\textbf {\bibinfo {volume} {107}},\ \bibinfo {pages} {140404} (\bibinfo {year} {2011})}\BibitemShut {NoStop}%
\bibitem [{\citenamefont {Enzian}\ \emph {et~al.}(2021)\citenamefont {Enzian}, \citenamefont {Price}, \citenamefont {Freisem}, \citenamefont {Nunn}, \citenamefont {Janousek}, \citenamefont {Buchler}, \citenamefont {Lam},\ and\ \citenamefont {Vanner}}]{Enzian2021}%
  \BibitemOpen
  \bibfield  {author} {\bibinfo {author} {\bibfnamefont {G.}~\bibnamefont {Enzian}}, \bibinfo {author} {\bibfnamefont {J.~J.}\ \bibnamefont {Price}}, \bibinfo {author} {\bibfnamefont {L.}~\bibnamefont {Freisem}}, \bibinfo {author} {\bibfnamefont {J.}~\bibnamefont {Nunn}}, \bibinfo {author} {\bibfnamefont {J.}~\bibnamefont {Janousek}}, \bibinfo {author} {\bibfnamefont {B.~C.}\ \bibnamefont {Buchler}}, \bibinfo {author} {\bibfnamefont {P.~K.}\ \bibnamefont {Lam}},\ and\ \bibinfo {author} {\bibfnamefont {M.~R.}\ \bibnamefont {Vanner}},\ }\bibfield  {title} {\bibinfo {title} {Single-phonon addition and subtraction to a mechanical thermal state},\ }\href@noop {} {\bibfield  {journal} {\bibinfo  {journal} {Phys. Rev. Lett.}\ }\textbf {\bibinfo {volume} {126}},\ \bibinfo {pages} {033601} (\bibinfo {year} {2021})}\BibitemShut {NoStop}%
\bibitem [{\citenamefont {Um}\ \emph {et~al.}(2016)\citenamefont {Um}, \citenamefont {Zhang}, \citenamefont {Lv}, \citenamefont {Lu}, \citenamefont {An}, \citenamefont {Zhang}, \citenamefont {Nha}, \citenamefont {Kim},\ and\ \citenamefont {Kim}}]{Um2016}%
  \BibitemOpen
  \bibfield  {author} {\bibinfo {author} {\bibfnamefont {M.}~\bibnamefont {Um}}, \bibinfo {author} {\bibfnamefont {J.}~\bibnamefont {Zhang}}, \bibinfo {author} {\bibfnamefont {D.}~\bibnamefont {Lv}}, \bibinfo {author} {\bibfnamefont {Y.}~\bibnamefont {Lu}}, \bibinfo {author} {\bibfnamefont {S.}~\bibnamefont {An}}, \bibinfo {author} {\bibfnamefont {J.-N.}\ \bibnamefont {Zhang}}, \bibinfo {author} {\bibfnamefont {H.}~\bibnamefont {Nha}}, \bibinfo {author} {\bibfnamefont {M.~S.}\ \bibnamefont {Kim}},\ and\ \bibinfo {author} {\bibfnamefont {K.}~\bibnamefont {Kim}},\ }\bibfield  {title} {\bibinfo {title} {Phonon arithmetic in a trapped ion system},\ }\href@noop {} {\bibfield  {journal} {\bibinfo  {journal} {Nature Comm.}\ }\textbf {\bibinfo {volume} {7}},\ \bibinfo {pages} {11410} (\bibinfo {year} {2016})}\BibitemShut {NoStop}%
\bibitem [{\citenamefont {Martienssen}\ and\ \citenamefont {Spiller}(1964)}]{Spiller1964}%
  \BibitemOpen
  \bibfield  {author} {\bibinfo {author} {\bibfnamefont {W.}~\bibnamefont {Martienssen}}\ and\ \bibinfo {author} {\bibfnamefont {E.}~\bibnamefont {Spiller}},\ }\bibfield  {title} {\bibinfo {title} {Coherence and fluctuations in light beams},\ }\href {https://doi.org/10.1119/1.1970023} {\bibfield  {journal} {\bibinfo  {journal} {American Journal of Physics}\ }\textbf {\bibinfo {volume} {32}},\ \bibinfo {pages} {919} (\bibinfo {year} {1964})}\BibitemShut {NoStop}%
\bibitem [{\citenamefont {Hlou\ifmmode~\check{s}\else \v{s}\fi{}ek}\ \emph {et~al.}(2024)\citenamefont {Hlou\ifmmode~\check{s}\else \v{s}\fi{}ek}, \citenamefont {Grygar}, \citenamefont {Dudka},\ and\ \citenamefont {Je\ifmmode~\check{z}\else \v{z}\fi{}ek}}]{Hlousek2024}%
  \BibitemOpen
  \bibfield  {author} {\bibinfo {author} {\bibfnamefont {J.}~\bibnamefont {Hlou\ifmmode~\check{s}\else \v{s}\fi{}ek}}, \bibinfo {author} {\bibfnamefont {J.}~\bibnamefont {Grygar}}, \bibinfo {author} {\bibfnamefont {M.}~\bibnamefont {Dudka}},\ and\ \bibinfo {author} {\bibfnamefont {M.}~\bibnamefont {Je\ifmmode~\check{z}\else \v{z}\fi{}ek}},\ }\bibfield  {title} {\bibinfo {title} {High-resolution coincidence counting system for large-scale photonics applications},\ }\href {https://doi.org/10.1103/PhysRevApplied.21.024023} {\bibfield  {journal} {\bibinfo  {journal} {Phys. Rev. Appl.}\ }\textbf {\bibinfo {volume} {21}},\ \bibinfo {pages} {024023} (\bibinfo {year} {2024})}\BibitemShut {NoStop}%
\bibitem [{\citenamefont {Sperling}\ \emph {et~al.}(2014)\citenamefont {Sperling}, \citenamefont {Vogel},\ and\ \citenamefont {Agarwal}}]{Sperling2014}%
  \BibitemOpen
  \bibfield  {author} {\bibinfo {author} {\bibfnamefont {J.}~\bibnamefont {Sperling}}, \bibinfo {author} {\bibfnamefont {W.}~\bibnamefont {Vogel}},\ and\ \bibinfo {author} {\bibfnamefont {G.~S.}\ \bibnamefont {Agarwal}},\ }\bibfield  {title} {\bibinfo {title} {Quantum state engineering by click counting},\ }\href {https://doi.org/10.1103/PhysRevA.89.043829} {\bibfield  {journal} {\bibinfo  {journal} {Phys. Rev. A}\ }\textbf {\bibinfo {volume} {89}},\ \bibinfo {pages} {043829} (\bibinfo {year} {2014})}\BibitemShut {NoStop}%
\bibitem [{git()}]{github}%
  \BibitemOpen
  \href {https://github.com/PepaHlousek/eme} {\bibinfo {title} {https://github.com/pepahlousek/eme}}\BibitemShut {NoStop}%
\end{thebibliography}%
 
\end{document}